\documentstyle[12pt]{article}
\def\lsim{\ \lower-0.4ex\hbox{$<$}\kern-0.80em\lower0.8ex\hbox{$\sim$}\ }
\def\gsim{\ \lower-0.4ex\hbox{$>$}\kern-0.80em\lower0.8ex\hbox{$\sim$}\ }
\begin{document}
\baselineskip=14pt
\vspace*{-1.0cm}

\begin{flushright} 
RESCEU-12/96, UTAP-241/96
\end{flushright}

\vspace*{0.3cm}
\baselineskip=24pt
\begin{center}
{\bf SPIN-DOWN OF NEUTRON STARS AND COMPOSITIONAL TRANSITIONS IN 
THE COLD CRUSTAL MATTER}
\end{center}
\vspace*{0.2cm}

\baselineskip=24pt
\begin{center}
Kei Iida$^{1}$ and Katsuhiko Sato$^{1,2}$
\vspace*{0.4cm}

\baselineskip=18pt
$^{1}$Department of Physics, School of Science, the University of Tokyo

7-3-1 Hongo, Bunkyo, Tokyo 113, Japan;

iida@utaphp1.phys.s.u-tokyo.ac.jp, sato@phys.s.u-tokyo.ac.jp

$^{2}$Research Center for the Early Universe, School of Science,
the University of Tokyo

7-3-1 Hongo, Bunkyo, Tokyo 113, Japan

\vspace*{0.7cm}
{\bf ABSTRACT}
\end{center}

\baselineskip=18pt

Transitions of nuclear compositions in the crust of a neutron star 
induced by stellar spin-down are evaluated at zero temperature.
We construct a compressible liquid-drop model for the energy of
nuclei immersed in a neutron gas, 
including pairing and shell correction terms,
in reference to the known properties of the ground state of matter
above neutron drip density, $4.3\times 10^{11}$ g cm$^{-3}$. 
Recent experimental values and extrapolations of nuclear masses
are used for a description of matter at densities below neutron drip. 
Changes in the pressure of matter in the crust
due to the stellar spin-down are calculated by taking 
into account the structure of the crust of
a slowly and uniformly rotating relativistic neutron star.
If the initial rotation period is $\sim$ ms,
these changes cause nuclei, initially being in the ground-state matter
above a mass density of about $3\times 10^{13}$ g $\rm{cm}^{-3}$,
to absorb neutrons in the equatorial region where the matter undergoes
compression, and to emit them in the vicinity of the rotation axis where
the matter undergoes decompression. 
Heat generation by these processes is found to have significant
effects on the thermal evolution of old neutron stars with low
magnetic fields; the surface emission predicted from this heating 
is compared with the $ROSAT$ observations of X-ray emission
from millisecond pulsars and is shown to be insufficient to explain
the observed X-ray luminosities.

\vspace*{0.5cm}
\noindent
{\it Subject headings:} dense matter ----- nuclear reactions,
nucleosynthesis, abundances ----- 
\hspace*{2.8cm}stars: interiors ----- stars: 
neutron ----- stars: rotation ----- X-rays: stars 

\baselineskip=18pt
\noindent
\begin{center}
{\bf 1.~ INTRODUCTION}
\end{center}
\vspace*{0.5cm}

A neutron star crust consists of a lattice of nuclei embedded in a
roughly uniform neutralizing background of electrons and, at
densities above neutron drip, in a sea of neutrons. 
So far, properties of the ground-state matter in the crust, 
such as the nuclei present and the equation of state, 
have been well investigated on the basis of laboratory data 
on nuclei and many-body calculations of the properties of
uniform nuclear matter (see, e.g., Pethick \& Ravenhall 1995).
If the crustal matter, initially in its ground state, is compressed or
decompressed, the matter is expected to deviate from nuclear
equilibrium; the constituents of matter, giving its
lowest energy state, now turn into another ones.  
It is expected, furthermore, that the matter being out of nuclear 
equilibrium undergoes nuclear processes such as electron captures,
$\beta-$decays, neutron absorption and emission, and nuclear fusion
and fission; 
these processes can release nuclear energy
to lower the total energy of the matter.

One of the astrophysical situations in which the pressure of
matter in the crust is changing
is in evolved neutron-star-black-hole binary systems;
there may occur the tidal disruption of neutron stars by black holes. 
Lattimer et al.\ (1977) have investigated the compositions of expanding,
initially cold, neutron star matter appropriate to these systems. 
It was noted that the compositions in the decompression bear
a resemblance to those of
hot r-process matter following a supernova explosion. Another situation is
when a neutron star accretes matter from a companion star. 
If matter of $\sim 0.01 M_{\odot}$ is accreted on to the surface
of the star, the crust is composed totally of this accreted matter.
Sato (1979)
and Haensel \& Zdunik (1990a) have examined the compositions
of cold matter initially of a 
density of about $10^{9}$ g cm$^{-3}$ being
compressed up to $\sim 10^{13}$ g cm$^{-3}$,
which we may encounter in accreting neutron stars. 
Their conclusion was that the charge
number of nuclei in the matter
decreases as a result of electron captures and that, 
once neutrons drip out of the nuclei, the mass number also decreases by
neutron emission. At the highest densities where Coulomb barrier
between the nuclei is sufficiently low, pycnonuclear fusion was found
to occur.

$ROSAT$ observations of X-ray emission from rotation-powered pulsars
give us important information on the surface emission from isolated
(not accreting) 
neutron stars of various ages (\"{O}gelman 1995). As the rotation is 
slowing down, the matter in the crust far away from the rotation axis is
expected to be compressed due to decrease in the centrifugal force; 
on the other hand, the matter in the polar region may be decompressed
due to increase in the length of the star along this axis. In this paper, 
we shall then 
investigate transitions of the nuclear compositions in the crust 
under such compression and decompression 
and effects of the release of nuclear energy
on the thermal evolution of neutron stars. 

During the time scale for the spin-down $\tau_{s}$,
which is $\gsim10$ yr as deduced from a magnetic dipole braking, 
neutrino cooling processes work so effectively that
the temperature of the crust is below $\sim10^{9}$ K
(see, e.g., Nomoto \& Tsuruta 1987). For such temperatures, thermal
energies are smaller than typical excitation energies 
of the system except in the outermost layers of the star 
where the thermal motion of electrons and ions affects the 
matter properties (Pethick \& Ravenhall 1995).
In describing matter in the crust of mass densities 
$\rho\gsim 10^{9}$ g cm$^{-3}$ considered below,
therefore, we assume its temperture to be zero. 
We also assume that this zero-temperature matter is in nuclear
equilibrium before the compression and decompression proceed
due to the spin-down.

Nuclear processes can occur in the cold crustal matter 
during the spin-down only when two conditions are satisfied. 
The first is that the time scale
for the processes must be less than $\tau_{s}$, 
and the second is that the Gibbs free 
energy of the system should be lowered by the processes.
We first consider these criteria for electron captures,
$\beta$-decays, and neutron absorption and emission.
The time scale for neutron processes occurring above neutron drip
density via the interactions of neutrons in continuum states
with nuclei (Lattimer et al.\ 1977) is negligibly small
compared with $\tau_{s}$. 
According to the estimates made by
Bisnovatyi-Kogan \& Seidov (1970) and Lattimer et al.\ (1977), 
the time scale for $\beta$-processes, dependent on the nuclear 
structure and on the electron Fermi energies, 
may possibly be smaller than $\tau_{s}$
in the density region of interest to us, 
although this is still uncertain for the neutron-rich nuclei 
present near and above neutron drip density.
The energy criterion for the occurrence of the processes considered
may be re-expressed as
the chemical potential of electrons
(dripped neutrons) being across the energy threshold which is given by 
the difference in nuclear masses between before and after 
$\beta$-processes (neutron processes). 
Generally, this energy threshold depends strongly on
$pairing$ $and$ $shell$ effects in the nuclei,
so that these effects should be
duly taken into account in describing the nuclei in the crust. 
Next, let us inquire when nuclear fusion and fission
can occur. The time scale for these
processes is sensitive to the nuclear charge $Z$,
since $Z$ determines the size of Coulomb barrier between the nuclei and
that of fission barrier (Lattimer et al.\ 1977; Sato 1979).
Energetically, fusion and fission reactions can take place if the
nuclear binding energy per nucleon becomes larger
after these reactions. For both conditions, the occurrence of fusion and
fission requires sufficiently small and large values of $Z$,
respectively, compared with those for the equilibrium nuclei.  
  
For the purpose of examining the chemical evolution of the
neutron star crust as the star spins down, we construct a compressible
liquid-drop model of nuclei in the inner crust (at densities
above neutron drip) based on that of Baym, Bethe, \& Pethick
(1971; hereafter referred to as BBP) and use a description of matter
in the outer crust (at densities below neutron drip) by Baym, Pethick,
\& Sutherland (1971; hereafter referred to as BPS)\@. 
For the matter above neutron drip, not only is
a nuclear surface energy term of BBP modified but also a 
shell correction term is added to the BBP expressions
in such a way that 
the resulting compositions of the ground-state matter 
are consistent with those
obtained by Negele \& Vautherin (1973; 
hereafter referred to as NV) in their Hartree-Fock calculations
of complete unit cells with an effective nucleon-nucleon interaction. 
A pairing correction term obtained by M\"{o}ller \& Nix (1992)
as an empirical formula is also added to our nuclear model; 
an isospin dependence of nuclear pairing gaps found for laboratory 
nuclei was incorporated into their formula. 
Recent experimental values of the nuclear masses
(Audi \& Wapstra 1993) and their extrapolations (M\"{o}ller et al.\ 1995)
are built in the BPS expressions for the energy of matter below neutron drip. 
The resultant equation of state of the ground-state matter in the
crust is found to be consistent with that of BPS\@. 
The next step is to calculate changes in the pressure of matter in the
crust due to the stellar spin-down. We first describe the evolution of
structure of the crust by using an equation of structure
of a slowly and uniformly rotating relativistic star 
advanced by Hartle (1967) and the equation of state of BPS
as well as by assuming a magnetic dipole spin evolution.
Displacements of matter in the crust during the spin-down 
are then analyzed in the framework of general relativity.
By comparing these displacements with the evolving pressure profile of
the crust, we evaluate the variations in the pressure of the matter.
Using the constructed nuclear model and the evaluated 
pressure variations
as well as recalling the energy and time-scale 
conditions for the occurrence of nuclear processes described in 
the preceding paragraph, 
we investigate nuclear processes occurring in the crust
during the spin-down, the resulting heat generation, and
its effects on the thermal evolution of neutron stars. 
 
In \S\ 2, we first construct a model for zero-temperature matter
in the crust, and we
thereby calculate the compositions and equation of state of the 
matter in nuclear equilibrium. In \S\ 3, a model for the crust of 
a neutron star that is rotating down
is described; changes in the pressure of the crustal 
matter induced by the spin-down are calculated therefrom. 
In \S\ 4, we evaluate transitions of the compositions 
of the zero-temperature matter,
undergoing the changes in its pressure as calculated in \S\ 3, 
from the equilibrium ones obtained in \S\ 2. 
In \S\ 5, a heating rate associated with these transitions is
estimated, and its relation to the $ROSAT$ observations of
millisecond pulsars is examined. 
Conclusions are finally given in \S\ 6. 

\vspace*{1.0cm}

\noindent
\begin{center}
{\bf 2.~ MODELS FOR THE COLD CRUSTAL MATTER}
\end{center}
\vspace*{0.5cm}

In this section we construct an expression for the Gibbs free energy 
of zero-tempera-ture matter in the neutron star crust. We then
investigate the equilibrium nuclei present and the equation of
state for these nuclei by minimizing the Gibbs free energy of the
system under charge neutrality.
We assume that the matter is composed of a single
species of nucleus with nucleon number $A$ and proton number $Z$ at
a given pressure $P$.

\vspace{1.0cm}
\noindent
\begin{center}
2.1.~{\it Inner Crust}
\end{center}
\vspace*{0.5cm}

A compressible liquid-drop model originally developed by BBP
is useful for describing higher-energy states of cold matter
in the inner crust on a footing equal to its ground state. BBP gave
the energy $W_{N}(A,Z,V_{N},n_{n})$ of a spherical nucleus of
volume $V_{N}$ immersed in the uniform neutron gas of number density 
$n_{n}$ as  
\begin{equation}
  W_{N}=[(1-x)m_{n}c^{2}+xm_{p}c^{2}+W(k,x)]A+W_{\rm Coul}(Z,V_{N})
        +W_{\rm surf}^{\rm BBP}(A,Z,V_{N},n_{n})\ .
\end{equation}
Here $m_{n}(m_{p})$ is the neutron (proton) rest mass, $W(k,x)$ is
the energy per nucleon for bulk nuclear matter of number density
$n=2k^{3}/3{\pi}^{2}=A/V_{N}$ and of proton fraction $x=Z/A$ 
[see equation (3.19) of BBP],
$W_{\rm Coul}(Z,V_{N})$ is the Coulomb energy of protons
distributed uniformly in the nucleus, 
\begin{equation}
  W_{\rm Coul}=\frac{3}{5}\frac{Z^{2}e^{2}}{r_{N}}
\end{equation}
with $r_{N}=(3V_{N}/4\pi)^{1/3}$,
and $W_{\rm surf}^{\rm BBP}(A,Z,V_{N},n_{n})$
is the nuclear surface energy,
\begin{equation}
  W_{\rm surf}^{\rm BBP}=\frac{\sigma[W(k_{n},0)-W(k,x)]}{{\omega}_{0}}
    {\left( 1-\frac{n_{n}}{n} \right)}^{2/3}A^{2/3}
\end{equation}
with $k_{n}=(3{\pi}^{2}n_{n}/2)^{1/3}$, $\sigma =21.0$ MeV, and
${\omega}_{0}=16.5$ MeV\@.   

In the ground state of matter above neutron drip density,
we take note of the nuclear charge
obtained by NV in their
Hartree-Fock (HF) calculations of complete unit cells. 
The $Z$ values calculated by BBP agree
fairly well with those by NV at mass densities $\rho$ between
$4.3\times 10^{11}$ and $10^{13}$ g $\rm{cm}^{-3}$, but significantly 
exceeds them with increasing density for $\rho \gsim 10^{13}$ g
$\rm{cm}^{-3}$ (see Figure 6 of NV)\@. 
Such behavior of $Z$ is due to overestimation of
the nuclear surface energy by BBP, which was clarified by
Ravenhall, Bennett, \& Pethick (1972; hereafter referred to as RBP) 
through comparison with their HF calculations of the surface energy
with a Skyrme interaction. We thus affix to $W_{\rm surf}^{\rm BBP}$
a coefficient $\lambda$ acting to reduce this surface energy:
\begin{equation}
  \lambda=\left\{ \begin{array}{ll}
           1\ ,
         & \mbox{$n_{n}=0$} \\
           \tanh \left( \frac {C_{1}}{\mu_{n}} \right)\ , 
         & \mbox{$n_{n}>0\ ,$} 
 \end{array} \right.
\end{equation}
where 
\begin{equation}
  \mu_{n}=W(k_{n},0)+\frac{1}{3}k_{n}
	\frac{\partial W(k_{n},0)}{\partial k_{n}}
\end{equation}
is the chemical potential of dripped neutrons (not including the rest
mass), and $C_{1}$ is the adjustable parameter to be determined below.
Hereafter we replace $W_{\rm surf}^{\rm BBP}$ in equation (1) by 
$W_{\rm surf}$ defined as $W_{\rm surf}=\lambda W_{\rm surf}^{\rm BBP}$ 
and refer to the equation thereby modified as equation (1)$'$. 

The nuclear shell effects, giving rise to greater stability of the nuclei
with neutron or proton magic numbers, also affect the charge number
of the equilibrium nuclei; these effects were 
appropriately taken into account by NV through a proton spin-orbit force. 
We then add to equation (1)$'$ a shell correction term that was originally
developed for isolated nuclei by Myers \& Swiatecki (1966) and
extended to nuclei in the dripped-neutron regime by Sato (1979):
\begin{eqnarray}
  W_{\rm shell} &=& 5.8 \left\{\frac{F(N,N_{i})\chi
    +F(Z,Z_{i})\chi+F(Z,Z_{i}')(1-\chi)}{(A/2)^{2/3}}
     \right.  \nonumber  \\ & & \hspace*{3cm}
    -0.26[(1-x)\chi+x]A^{1/3} 
    \left.\frac{}{}\right\}\ {\rm MeV}
\end{eqnarray}
with
\begin{equation}
  F(X,M_{i})=\frac{3}{5} \left[(M_{i}^{5/3}-M_{i-1}^{5/3})
             \frac{X-M_{i-1}}{M_{i}-M_{i-1}}-(X^{5/3}-M_{i-1}^{5/3})  
             \right]\ ,~ M_{i-1}\leq X\leq M_{i}\ ,
\end{equation}
\begin{equation}
  \chi =\frac{1}{1+\exp \left(\frac{\mu_{n}'}{C_{2}} \right)}\ ,
\end{equation}
where $N=A-Z$ is the number of neutrons in the nucleus, 
$N_{i}=\{2,8,14,28,50,82$,$126$, $184$,$\cdot \cdot \cdot \}$ is the
set of neutron magic numbers for isolated nuclei,
$Z_{i}=\{2,8$,$14$,$28$,$50$,$82$,$114$,$\cdot$ $\cdot \cdot \}$ is
the set of proton magic numbers for isolated nuclei,
$Z_{i}'=\{2,8$,$20$,$28$,$40$,$50$,$82$,$114$,$\cdot$
$\cdot \cdot \}$ is the set of proton magic numbers for nuclei in the
external neutron gas found by NV, and
\begin{equation}
  \mu_{n}'=\frac{\partial}{\partial A}[AW(k,x)
           +W_{\rm surf}(A,Z,V_{N},n_{n})]_{Z,V_{N},n_{n}}
\end{equation}
is roughly the chemical potential of neutrons in the nucleus.
The factor $\chi$ ensures that equation (6) reproduces the 
empirical shell formula of Myers \& Swiatecki (1966) for isolated nuclei, 
and that the dripped neutrons smear the neutron shells and 
turn the proton magic numbers $Z_{i}$ into $Z_{i}'$ (see Sato 1979).
 
The parameter $C_{2}$ has been determined together with  $C_{1}$ 
in such a way as to reproduce the $Z$ values of 
the ground-state matter calculated by NV at eleven mass densities, 
so that $C_{1}=5.5$ MeV and $C_{2}=0.1$ MeV\@. Strictly speaking, 
we have obtained $Z=40$ in place of $Z=32$ (NV) at the highest 
density. In this determination we have evaluated the
values of $Z$ by minimizing the Gibbs free energy of the
system under charge neutrality, as will be discussed in \S\ 2.3. 

In Figure 1 we have compared the resultant surface tension,
$E_{s}=W_{\rm surf}/4\pi r_{N}^{2}$, with the BBP and RBP results.
The present values approach the HF values of RBP with
increasing density (decreasing proton fraction) for $\rho\gsim 10^{13}$
g cm$^{-3}$ $(x\lsim 0.26)$ and reproduce the BBP values in the
density range $4.3\times 10^{11}$ g cm$^{-3}$ $\leq\rho\lsim 10^{13}$
g cm$^{-3}$ $(0.26\lsim x\leq 0.31)$. An improvement of the present
surface energy over that of BBP has been thus clarified. 

We next introduce the nuclear pairing effects, stabilizing the 
nuclei with even neutrons and even protons relative to
the other types of nuclei, 
into the nuclear model
hitherto determined by adding a pairing correction term obtained 
empirically by M\"{o}ller \& Nix (1992) in their macroscopic model:
\begin{equation}
  W_{\rm pair}=\left\{ \begin{array}{llll}
           0
         & \mbox{(even-neutron, even-proton)} \\
           \frac{r_{\rm MN}B_{s}}{N^{1/3}}{\rm{e}}^{-t_{\rm MN}\delta^{2}} 
         & \mbox{(odd-neutron, even-proton)} \\
           \frac{r_{\rm MN}B_{s}}{Z^{1/3}}{\rm{e}}^{-t_{\rm MN}\delta^{2}} 
         & \mbox{(even-neutron, odd-proton)} \\
           \frac{r_{\rm MN}B_{s}}{N^{1/3}}{\rm{e}}^{-t_{\rm MN}\delta^{2}}+ 
           \frac{r_{\rm MN}B_{s}}{Z^{1/3}}{\rm{e}}^{-t_{\rm MN}\delta^{2}}- 
           \frac{h_{\rm MN}}{A^{2/3}B_{s}} 	
         & \mbox{(odd-neutron, odd-proton)} \\
 \end{array} \right.
\end{equation}
with $h_{\rm MN}=6.6$ MeV, $r_{\rm MN}=5.55$ MeV, and $t_{\rm MN}=6.07$.
Here $\delta=1-2x$ is the relative neutron excess, and 
$B_{s}$ is the surface area of a deformed nucleus divided by
that of a spherical one. We simply set $B_{s}=1$. It is to be noted
that they determined $W_{\rm pair}$ from experimental 
mass differences in the isospin range $-0.1<\delta<0.25$, and that
we shall extrapolate $W_{\rm pair}$
into the dripped-neutron regime $(0.4\lsim \delta\lsim 0.7)$.
We also note that this pairing correction term, 
being measured relative to the even-even nuclei, 
produces no change in
the charge number of the equilibrium nuclei.   
 
We finally obtain the Gibbs free energy of the system above neutron drip
per nucleon, as BBP did, according to
\begin{equation}
  g=\frac{E_{\rm tot}+P}{n_{N}A_{\rm cell}}
\end{equation}
with 
\begin{equation}
  E_{\rm tot}=n_{N}m_{N}(A,Z,V_{N},n_{N},n_{n})+{\varepsilon}_{e}(n_{e})
    +(1-n_{N}V_{N}){\varepsilon}_{n}(n_{n})\ ,	
\end{equation}
where $E_{\rm tot}$ is the total energy per unit volume, which is in 
turn related to $\rho$ as $E_{\rm tot}=\rho c^{2}$, and $n_{N}$ is 
the number of nuclei per unit volume. Here,
\begin{equation}
  m_{N}=W_{N}(A,Z,V_{N},n_{n})+W_{L}(Z,V_{N},n_{N})
       +W_{\rm shell}(A,Z,V_{N},n_{n})+W_{\rm pair}(A,Z) 
\end{equation}
is the sum of the mass of a nucleus given by equations (1)$'$, (6), 
and (10) and the lattice energy in the
Wigner-Seitz approximation,
\begin{equation}
  W_{L}=-\frac{9}{10}\frac{Z^{2}e^{2}}{r_{c}}
        \left(1-\frac{1}{3}\frac{r_{N}^{2}}{r_{c}^{2}} \right)
\end{equation}
with $r_{c}=(3/4\pi n_{N})^{1/3}$, 
\begin{equation}
  {\varepsilon}_{e}=\frac{m_{e}^{4}c^{5}}{8{\pi}^{2}{\hbar}^{3}}
                   \{(2t^{2}+1)t(t^{2}+1)^{1/2}
                    -\ln [t+(t^{2}+1)^{1/2}]\} 
\end{equation}
with $t=\hbar(3{\pi}^{2}n_{e})^{1/3}/m_{e}c$ is the energy of the
free electron gas per unit volume,
\begin{equation}
  {\varepsilon}_{n}=n_{n}[W(k_{n},0)+m_{n}c^{2}]
\end{equation}
is the energy of the external neutron gas per unit volume, and
\begin{equation}
  A_{\rm cell}=A+n_{n}\left(\frac{1}{n_{N}}-V_{N}\right)
\end{equation}
is the nucleon number in a spherical Wigner-Seitz cell, which is equal
to the ratio of the number density of baryons $n_{b}$ to $n_{N}$. 

\vspace{1.0cm}
\noindent
\begin{center}
2.2.~{\it Outer Crust}
\end{center}
\vspace*{0.5cm}

To construct a model of matter in the outer crust, we follow
a line of arguments of BPS and set the Gibbs free
energy of the system per nucleon as
\begin{equation}
  g=\frac{E_{\rm tot}+P}{n_{N}A}
\end{equation}
with
\begin{equation}
  E_{\rm tot}=n_{N}m_{N}(A,Z,n_{N})+{\varepsilon}_{e}(n_{e})\ ,
\end{equation}
where ${\varepsilon}_{e}$ is given by equation (15), and
\begin{equation}
  m_{N}=M_{N}(A,Z)+W_{L}(Z,n_{N})
\end{equation}
is the sum of the mass of a nucleus $M_{N}(A,Z)$ and
the energy of a bcc lattice,
\begin{equation}
  W_{L}=-0.895929\frac{Z^{2}e^{2}}{r_{c}}\ .
\end{equation}
Here the nuclear mass, not having the atomic-electron binding 
energy subtracted out, is related to the atomic mass $M_{A}(A,Z)$
as $M_{N}(A,Z)=M_{A}(A,Z)-Zm_{e}c^{2}$.
The values of $M_{A}(A,Z)$ used here are taken from the
experimental mass table of Audi \& Wapstra (1993), when present.
The masses of the remaining atoms are taken from the droplet 
(FRDM) mass table of M\"{o}ller et al.\ (1995). 
We note that the matter model 
described above is the same as that of Haensel \& Pichon (1994)
except some minor differences concerning the electronic contribution
to binding energies, the electron exchange energies, 
and the electron-screening effects.  
 
\vspace{1.0cm}
\noindent
\begin{center}
2.3.~{\it Ground-State Properties}
\end{center}
\vspace*{0.5cm}

It is necessary to know the ground-state properties of matter in
the crust for determining the initial model for the equation of state 
and the nuclear compositions. 
The equilibrium conditions for the matter are obtained
by minimizing the Gibbs free energy per nucleon, $g$, given by equations
(11) and (18) at fixed $P$ under charge neutrality,
\begin{equation}
  n_{e}=Zn_{N}\ .
\end{equation}
For the matter in the inner crust, as investigated by BBP,
the conditions for determining $n_{N}$ and $V_{N}$  
at fixed $P$, $A$, $Z$, and $n_{n}$ are the pressure equilibrium
condition,
\begin{equation}
 -\left.\frac{\partial m_{N}}{\partial V_{N}}\right|_{A,Z,n_{N},n_{n}}
  =P^{\rm (G)}\equiv\frac{1}{3}n_{n}k_{n}\frac{\partial W(k_{n},0)}
   {\partial k_{n}}+\frac{n_{n}n_{N}}{1-V_{N}n_{N}}
  \left.\frac{\partial m_{N}}{\partial n_{n}}\right|_{A,Z,V_{N},n_{N}}
\end{equation}
and the thermodynamic relation between $P$ and $n_{b}$,
\begin{equation}
  P=P^{\rm (G)}+n_{e}^{2}\frac{\partial}{\partial n_{e}}
   \left(\frac{\varepsilon_{e}}{n_{e}}\right)+n_{N}^{2}
   \left.\frac{\partial W_{L}}{\partial n_{N}}\right|_{Z,V_{N}}\ .
\end{equation} 
The remaining conditions for determining $A$, $Z$, and $n_{n}$ at
fixed $P$, $V_{N}$, and $n_{N}$ are the optimization of $A$ in a 
nucleus, 
\begin{equation}
  S_{A+1}<\frac{\partial}{\partial A}\left(\frac{W_{\rm Coul}+W_{\rm surf}
          +W_{L}}{A}\right)_{x,n_{N}A,n_{N}V_{N}, n_{n}}<S_{A}
\end{equation} 
with 
\begin{eqnarray}
  S_{A} &=& \frac{W_{\rm pair}(A-1,Z)+W_{\rm shell}
        \left(A-1,Z,\frac{A-1}{A}V_{N},
        \frac{A}{A-1}n_{N},n_{n}\right)}{A-1}
  \nonumber \\ & &
      -\frac{W_{\rm pair}(A,Z)+W_{\rm shell}(A,Z,V_{N},n_{N},n_{n})}{A}\ ,
\end{eqnarray} 
the $\beta-$stability condition,
\begin{equation}
  W_{Z}<\mu_{e}<W_{Z-1}
\end{equation} 
with
\begin{equation}
  W_{Z}=m_{N}(A,Z,V_{N},n_{N},n_{n})-m_{N}(A,Z+1,V_{N},n_{N},n_{n})\ ,
\end{equation} 
where $\mu_{e}=\partial {\varepsilon}_{e}/\partial n_{e}$ 
is the electron chemical potential including the rest 
mass,
and the equilibrium of the dripped neutrons with the neutrons
in a nucleus,   
\begin{equation}
  D_{A}<\mu_{n}+m_{n}c^{2}<D_{A+1}
\end{equation} 
with
\begin{equation}
  D_{A}=m_{N}(A,Z,V_{N},n_{N},n_{n})
  -m_{N}\left(A-1,Z,V_{N},n_{N},n_{n}+\frac{n_{N}}{1-n_{N}V_{N}}
  \right)\ .
\end{equation}
In the conditions (27) and (29), the quantities $W_{Z}$, $W_{Z-1}$, 
$D_{A}$, and $D_{A+1}$ correspond to the energy thresholds 
necessary for $\beta-$decay, electron capture, neutron emission, 
and neutron absorption, respectively, to occur in a cell containing
the nucleus $(A,Z)$.  
We note that inequalities (25), (27), and (29) are only necessary
conditions for determining the compositions of the
ground-state matter; hence, we should find the parameters $A$,
$Z$, and $n_{n}$, which give the minimum value of $g$, 
among those satisfying these inequalities.
 
For the matter in the outer crust, a complete description 
of the minimization of $g$ is given by BPS\@. 
In this case, the equilibrium conditions (24) and (27) 
remain useful if we set $P^{\rm (G)}=0$ and eliminate the 
parameters $V_{N}$ and $n_{n}$.

The neutron drip point for the ground-state matter has been
calculated by minimizing $g$ given by equation (18) and 
finding the case in which $g=m_{n}c^{2}$, as evaluated by BPS\@. 
The result indicates that a neutron begins to drip 
out of $^{118}$Kr at $\rho =4.34\times 10^{11}$ g $\rm{cm}^{-3}$,
$P=4.93\times 10^{-4}$ MeV $\rm{fm}^{-3}$, and $\mu_{e}=26.21$ MeV\@.
The agreement with the results of BPS and of Haensel \& Pichon (1994)
has been thus confirmed.

Using the equilibrium conditions discussed above,
we have obtained the equilibrium nuclear compositions 
for eleven cases of densities below neutron drip
as well as for eleven cases of densities above neutron drip,
which were chosen by NV, 
and determined 
the equation of state for these compositions; the result has been
listed in Table 1.
We have confirmed that the calculated nuclear compositions (including
the chemical potentials $\mu_{e}$ and $\mu_{n}$) are consistent 
with those of Haensel \& Pichon (1994) in the outer crust as well as
those of NV in the inner crust.
The equilibrium nuclear compositions listed in Table 1 will be
used as the initial model for the nuclear compositions 
in \S\ 4.
As depicted in Figure 2, the equation of state in the
present calculations agrees graphically with the equation of state
of BPS which includes that of BBP in the dripped-neutron regime. 
We shall thus use the equation of state of BPS as that consistent
with the present result for estimating the structure of the crust
in \S\ 3.

\vspace*{1.0cm}

\noindent
\begin{center}
{\bf 3.~ SPIN-DOWN AND CHANGES IN PRESSURE OF THE CRUSTAL MATTER}
\end{center}
\vspace*{0.5cm}

In this section we first construct a model for the crust of a 
zero-temperature neutron star rotating at a given angular velocity. 
By using a magnetic dipole approximation for the rotational evolution,
the evolution of structure of the crust is determined. We
finally estimate the displacements of elements of matter in the crust
during the spin-down and then the corresponding changes in their
pressures. We use units in which $c=G=1$ in \S\ 3.

\vspace{1.0cm}
\noindent
\begin{center}
3.1.~{\it Model for the Crust of a Rotating Neutron Star}
\end{center}
\vspace*{0.5cm}

As a first step towards the estimates of 
the changing pressures of matter in the crust during the spin-down,
we consider a pressure profile of the crust of
a neutron star that is in hydrostatic equilibrium
and is rotating at a uniform angular velocity
$\Omega =2\pi/P_{\rm rot}$, where $P_{\rm rot}$ is the rotation period 
as seen by a distant observer. 
In a non-rotating equilibrium configuration, we assume the stellar
mass $M$ and radius $R_{s}$ to be $1.4$ $M_{\odot}$ and $10$ km. 
We fix the total number of baryons in the star for arbitrary $\Omega$. 
General relativistic effects on 
the crust model may be deduced from the following quantity 
included in the redshift factor at the surface of the non-rotating star:
\begin{equation}
  \frac{2GM}{R_{s}c^{2}}\sim 0.4\ .
\end{equation}
A fully general relativistic treatment of the configuration of the crust
is thus required. If the star is rotating with $P_{\rm rot}=1$ ms,
being similar to the shortest period $1.6$ ms among those observed from
pulsars, the ratio between the typical magnitude of the rotational 
energy of the star and that of its gravitational potential is roughly
\begin{equation}
  \frac{R_{s}^{3}\Omega^{2}}{2GM}\sim 0.1\ .
\end{equation}
We can then consider the rotation as a small perturbation on a 
non-rotating configuration. Consequently, we find that 
general-relativistic equations of structure of 
slowly and rigidly rotating stars 
obtained by Hartle (1967) exactly up to order ${\Omega}^{2}$ 
are useful for our purpose of determining the pressure profile
of the crust of a rotating neutron star.

Such a pressure profile may be obtained by determining the
transformation of a surface of
constant pressure in the equilibrium configuration of the crust
of a neutron star rotating at a given $\Omega$ from a spherical 
one of radius $R$ in the non-rotating equilibrium configuration
subject to conservation of the number of baryons in the star. We denote
the position of the surface of constant pressure
in the rotating configuration by the
ordinary polar coordinates $r$ and $\theta$ in which the polar axis is 
taken to be the rotation axis. At a given $\theta$, the radius 
of the surface may be written as (Hartle 1967)
\begin{equation}
  r=R+\xi(R,\theta)+{\rm{O}}({\Omega}^{4})\ ,
\end{equation}
where 
\begin{equation}
  \xi(R,\theta)={\xi}_{0}(R)+{\xi}_{2}(R)P_{2}(\cos \theta)
\end{equation}
is the displacement of order $\Omega^{2}$ from its non-rotating
position. Here ${\xi}_{0}(R)$ denotes the change in the mean radius
of the surface, and ${\xi}_{2}(R)P_{2}(\cos \theta)$ represents
the quadrupolar deformation of the surface, 
where $P_{2}(\cos \theta)$ is the
second-order Legendre polynomial. For the present purpose, we should
then evaluate the radius $R$ of the surface of constant pressure in 
the configuration of the non-rotating neutron star crust and 
its displacement $\xi(R,\theta)$ due to the rotation.
  
We first calculate the pressure profile, $P(R)$, of the crust
of a non-rotating neutron star by using the equation of state
of BPS in the Tolman-Oppenheimer-Volkoff
(TOV) equations (see Shapiro \& Teukolsky 1983):
\begin{equation}
  \frac{dP(R)}{dR}=-\frac{[\rho(R)+P(R)][m(R)+4\pi R^{3}P(R)]}
                         {R^{2}}{\rm{e}}^{-\nu(R)}\ ,
\end{equation}
where  
\begin{equation}
  {\rm{e}}^{\nu(R)}=1-\frac{2m(R)}{R}
\end{equation}
with
\begin{equation}
  m(R)=M-\int_{R}^{R_{s}}dr 4\pi r^{2}\rho(r) 
\end{equation}
is the redshift factor.
The pressure profile $P(R)$ 
obtained by solving equation (35) inward from the surface of the star
is shown in Figure 3,
together with the density profile $n_{b}(R)$. 
Here we have assumed that the crust dissolves into
uniform nuclear matter at a density of $\approx 0.1$ ${\rm{fm}}^{-3}$, 
as suggested by Lorenz, Ravenhall, \& Pethick (1993).

We proceed to estimate the displacement, $\xi(R,\theta)$,
of the surface of constant
pressure in the crust from the non-rotating one due to the rotation,
using the results for this displacement obtained by Hartle (1967) and
Hartle \& Thorn (1968). Since the thickness of the crust is small
compared with $R_{s}$ as depicted in Figure 3, 
it can be assumed that 
$\xi(R,\theta)\approx\xi(R_{s},\theta)$ throughout the crust. 
To calculate the displacement of the surface of the star 
$\xi(R_{s},\theta)$, we note the solutions of Einstein's field
equations for a perfect fluid in the gravitational field 
of the rotating configuration that is characterized
by the metric having the following form
up to order $\Omega^{2}$ (Hartle 1967):
\begin{eqnarray}
  ds^{2} &=& -{\rm{e}}^{\nu(r)}
         \{1+2[h_{0}(r)+h_{2}(r)P_{2}(\cos \theta)]\}(dx^{0})^{2}
         \nonumber \\ & &
         +{\rm{e}}^{-\nu(r)}
         \left\{1+\frac{2[m_{0}(r)+m_{2}(r)P_{2}(\cos \theta)]}
         {r-2m(r)}\right\}dr^{2} 
         \nonumber \\ & &
         +r^{2}\{1+2[v_{2}(r)-h_{2}(r)]P_{2}(\cos \theta)\}
         \{d{\theta}^{2}+\sin^{2}\theta[d\phi-\omega(r)dx^{0}]^{2}\}\ ,
\end{eqnarray}
where $\nu(r)$ and $m(r)$ are given by equations (36) and (37).
At the surface of the star, the solutions for the first- and second-order
quantities contained in equation (38) are obtained by Hartle (1967) as
\begin{equation}
  \omega(R_{s})=\frac{2J}{R_{s}^{3}}
\end{equation}
\begin{equation}
  h_{0}(R_{s})=-\frac{\delta M}{R_{s}\left(1-\frac{2M}{R_{s}} \right)}
               +\frac{J^{2}}{R_{s}^{4}\left(1-\frac{2M}{R_{s}} \right)}
\end{equation}
\begin{equation}
  h_{2}(R_{s})=\frac{J^{2}}{MR_{s}^{3}}\left(1+\frac{M}{R_{s}} \right)
               +\frac{5}{8M^{3}}\left(Q-\frac{J^{2}}{M} \right)
               Q_{2}^{\ 2}\left(\frac{R_{s}}{M}-1 \right)
\end{equation}
\begin{equation}
  m_{0}(R_{s})=\delta M-\frac{J^{2}}{R_{s}^{3}}
\end{equation}
\begin{equation}
  m_{2}(R_{s})=(R_{s}-2M)\left[-h_{2}(R_{s})
            +\frac{6J^{2}}{R_{s}^{4}}\right]
\end{equation}
\begin{equation}
  v_{2}(R_{s})=-\frac{J^{2}}{R_{s}^{4}}+\frac{5}{4M^{3}}
     \frac{MQ-J^{2}}{\sqrt{R_{s}(R_{s}-2M)}}
     Q_{2}^{\ 1}\left(\frac{R_{s}}{M}-1 \right)\ ,
\end{equation}
where $J$ is the total angular momentum of the star related to
the moment of inertia $I$ of the star as $J=I\Omega$, $\delta M$ is
the change in mass of the star from its
non-rotating value, $Q$ is the quadrupole moment of the star, 
and $Q_{2}^{\ m}$ is the associated Legendre polynomial of the
second kind. At fixed baryon number, $\delta M$ is given by
(Hartle 1970)
\begin{equation}
  \delta M=\frac{1}{2}I\Omega^{2}\ .
\end{equation}
The quantity of $\xi_{2}(R_{s})$ can be derived from the integral of the
equation  of hydrostatic equilibrium for the rotating configuration
[see equations (28), (89), and (91) of Hartle (1967)] as
\begin{equation}
  \xi_{2}(R_{s})=-\left\{\frac{1}{3}\frac{R_{s}^{2}
            [\Omega-\omega(R_{s})]^{2}}{1-\frac{2M}{R_{s}}}
            +h_{2}(R_{s}) \right\}\frac{R_{s}^{2}}{M}
            \left(1-\frac{2M}{R_{s}}\right)\ .
\end{equation}

The parameters which still remain to be determined are $I$, $Q$, and
$\xi_{0}(R_{s})$. These values, dependent on the properties of neutron
star matter, have been taken on the basis of the results for the rotating
configuration at a stellar mass of about $1.4$ $M_{\odot}$ 
obtained by Hartle \&
Thorn (1968), using the $V_{\gamma}$ equation of state that is hard
rather than soft, as
\begin{equation}
  I=10^{45}\ {\rm g}\ {\rm cm}^{-3}
\end{equation}
\begin{equation}
  Q=0.04\left(\frac{\Omega}{\Omega_{0}}\right)^{2}MR_{s}^{2}
\end{equation}
\begin{equation}
  \xi_{0}(R_{s})=0.06\left(\frac{\Omega}{\Omega_{0}}\right)^{2}R_{s}\ ,
\end{equation}
where $\Omega_{0}$ is the angular velocity at $P_{\rm rot}=1$ ms.
Here we have taken into account differences 
between the rotations at fixed central density and 
the rotations at fixed baryon number, 
the former being assumed by Hartle \& Thorn (1968).
We have thus obtained
\begin{equation}
  \frac{\xi(R_{s},\theta)}{R_{s}}
     =(-0.039+0.149\sin^{2}\theta)
      \left(\frac{\Omega}{\Omega_{0}}\right)^{2}\ .
\end{equation}
Equation (50), as well as the solutions of the TOV equations obtained by
using the equation of state of BPS, gives the present model for the crust
of a neutron star rotating at a given $\Omega$.

In Figure 4 we have drawn the surface of the star derived from equation (50)
at $\Omega=\Omega_{0}$ together with the surface of radius $R_{s}$
at $\Omega=0$. We observe in this comparison
that the rotation acts to shorten (enlarge) the surface of the star in
the polar (equatorial) region.

\newpage

\noindent
\begin{center}
3.2.~{\it Rotational and Structural Evolution}
\end{center}
\vspace*{0.5cm}

We assume that the evolution of the rotation of the star is determined 
by the usual magnetic-dipole model for pulsars (see Shapiro \&
Teukolsky 1983) as
\begin{equation}
  \Omega(t)=\frac{\Omega(t=0)}{\sqrt{1+
  \frac{\Omega(t=0)^{2}B^{2}R_{s}^{6}}{3Ic^{3}}t}}\ .
\end{equation}
Here $B$ is the constant magnetic-dipole field at the magnetic pole
of the star which is taken to be on the equator,
and $t$ denotes the age of the star.
The constant $B$ is then expressed in terms of
the rotation period $P_{\rm rot}$ and its derivative 
${\dot P}_{\rm rot}$ as 
$B\approx\sqrt{10^{39}P_{\rm rot}{\dot P}_{\rm rot}}$;
the values of $B$ that are determined by the quantities
$P_{\rm rot}$ and ${\dot P}_{\rm rot}$ observed from pulsars  
range from $\sim10^{8}$ G to $\sim10^{12}$ G 
(Taylor, Manchester, \& Lyne 1993). 
By using equation (51), the time scale for the spin-down 
may be estimated as
$\tau_{s}\sim 6\times 10^{5}[10^{10} {\rm G}/B({\rm G})]^{2}
[\Omega_{0}/\Omega(t=0)]^{2}$ yr. 

Let us now ask how the structure of the crust is varying as the
rotation of the star is slowing down according to equation (51). 
We assume that the total number of baryons in the star  
remains constant during the spin-down, 
and that the star is in hydostatic equilibrium and
is rotating rigidly at $t=0$.
Since the period derivative ${\dot P}_{\rm rot}$ observed from
pulsars is extremely small (typically
$10^{-15}$ s ${\rm s}^{-1}$), the configuration of the star can be considered
to maintain its hydrostatic equilibrium and rigid rotation. Then, 
the displacement of the surface  of the star given by equation
(50) as well as the perturbative quantities given by equations (39)--(44)
may be redetermined at a given $t$
as $\xi(R_{s},\theta,t)$, $\omega(R_{s},t)$, etc.,
by substituting $\Omega(t)$ 
into $\Omega$ included in each equation. 
In addition, we neglect the variation in the equation of state
of matter in the crust during the spin-down,
an assumption to be justified by calculating
the equation of state of the matter
deviating from its ground state in \S\ 4.
The evolution of structure of the crust is thus described
in terms of the quantity $\xi(R_{s},\theta,t)$ 
and of the pressure profile of the
non-rotating configuration obtained by using the
equation of state of BPS\@.

\vspace{1.0cm}
\noindent
\begin{center}
3.3.~{\it Changes in the Pressure of the Crustal Matter}
\end{center}
\vspace*{0.5cm}

For our main purpose of investigating transitions of the
compositions of the cold crustal matter during the spin-down, it is
essential to calculate the Lagrangian changes in pressure associated
with elements of matter in the crust.
Such Lagrangian changes in pressure may be obtained by combining
the displacements of matter elements from their initial positions
with the Eulerian changes in pressure obtained in \S\ 3.2.  
For convenience, we denote 
the initial position of a matter element by the depth 
$\Delta r$, the distance measured from the surface 
of the star along the radial line,
and by the polar angle $\theta$. 
We then express the position of this element at a given $t$ 
by the depth
$\Delta r+\eta(\Delta r,\theta,t)$ and the polar angle 
$\theta+\alpha(\Delta r,\theta,t)$, as shown in Figure 5. 
The quantities
$\eta$ and $\alpha$ correspond to the displacements in the 
radial and angular directions between the element and an observer
who is initially located at the same point as the element
and is comoving with the surface of constant pressure in the
radial direction. 
    
The direction in which elements of matter in the crust move as 
the star is spinning down gives the relation between the quantities
$\eta$ and $\alpha$. 
Let us assume that the direction of movement of
an element present at $(r,\theta,t)$ is parallel to the sum of
the gravitational and centrifugal forces acting on the element.
In the hydrostatic equilibrium configuration considered here,
these forces are in balance with the pressure gradient.
As can be seen from the radius of the surface of constant pressure
given by equation (33), the $\theta$ component of the pressure
gradient is of order $\Omega^{2}$, whereas the $r$ component
is of zeroth order. 
We thus find $\alpha$ to be of order $\Omega^{4}$,
by comparison with $\eta$ of order 
$\Omega^{2}$. Consequently, the quantity $\alpha$ can be neglected 
within the framework of the present perturbative treatment.

To find the quantity $\eta(\Delta r,\theta,t)$, we note that each 
element of matter in the crust is to be identified by counting
the number of baryons inward from the surface of the star along
the radial line. This identification
may be made by following the baryon number conservation law
(see Misner \& Sharp 1964) according to
\begin{eqnarray}
 & & \int_{R_{s}-\Delta r}^{R_{s}}dR \sin\theta
  \left\{r^{2}{\rm e}^{-\frac{\nu(r)}{2}}[1+\Delta(r,\theta,t=0)]
  \right\}_{r=R+\xi(R_{s},\theta,t=0)}n_{b}(R)
 \nonumber \\ & &
  =\int_{R_{s}-\Delta r-\eta(\Delta r,\theta,t)}^{R_{s}}dR
   \sin\theta
  \left\{r^{2}{\rm e}^{-\frac{\nu(r)}{2}}[1+\Delta(r,\theta,t)]
  \right\}_{r=R+\xi(R_{s},\theta,t)}n_{b}(R)\ , 
\end{eqnarray}
where $n_{b}(R)$ is the density profile of the non-rotating
configuration shown in Figure 3, and
\begin{equation}
  \Delta(r,\theta,t)=\frac{(\sqrt{-{\sl g}}u^{0})_{r,\theta,t}}
  {r^{2}{\rm e}^{-\frac{\nu(r)}{2}}\sin\theta}-1
\end{equation}
is the factor of order $\Omega^{2}$
determined by the determinant ${\sl g}$ of the metric of
equation (38) and by the time component of the 4-velocity of the
matter,
\begin{eqnarray}
  u^{0}(r,\theta,t) &=&
   {\rm e}^{-\frac{\nu(r)}{2}}\left\{1+\frac{1}{2}r^{2}
   {\rm e}^{-\nu(r)}[\Omega(t)-\omega(r,t)]^{2}\sin^{2}\theta
   -h_{0}(r,t)-h_{2}(r,t)P_{2}(\cos \theta)\right\}
    \nonumber \\ & & 
   +{\rm O}(\Omega^{4})\ .  
\end{eqnarray}
We then set  
$\Delta(r,\theta,t)\approx\Delta(R_{s},\theta,t)$ in equation (52),
by noting the correspondence with the approximation,
$\xi(R,\theta,t)\approx\xi(R_{s},\theta,t)$, available
for thin crusts as discussed in \S\ 3.1. The values of
$\Delta(R_{s},\theta,t)$ can be calculated from
equations (36), (38), (39), (41)--(44), (54) as
\begin{eqnarray}
  \Delta(R_{s},\theta,t) &=& \frac{1}{2}{\rm e}^{-\nu(R_{s})}R_{s}^{2}
  \{[\Omega(t)-\omega(R_{s},t)]^{2}-\omega(R_{s},t)^{2}\}\sin^{2}\theta
    \nonumber \\ & &
  +2[v_{2}(R_{s},t)-h_{2}(R_{s},t)]P_{2}(\cos\theta)
  +\frac{m_{0}(R_{s},t)+m_{2}(R_{s},t)P_{2}(\cos\theta)}{R_{s}-2M}
    \nonumber \\ &=&
  (-0.043+0.094\sin^{2}\theta)\left
  [\frac{\Omega(t)}{\Omega_{0}}\right]^{2}\ .
\end{eqnarray}
We have thus obtained the quantity $\eta(\Delta r,\theta,t)$ by
retaining only the terms of order $\Omega^{2}$ in equation (52) as
\begin{eqnarray}
  \eta(\Delta r,\theta,t) &=& \left[\frac{{\rm e}^{\frac{\nu(R)}{2}}}
  {R^{2}n_{b}(R)}\right]_{R=R_{s}-\Delta r}
  \nonumber\\ & & \times
  \left\{[\Delta(R_{s},\theta,t=0)-\Delta(R_{s},\theta,t)]
  \int_{R_{s}-\Delta r}^{R_{s}}dR R^{2}{\rm e}^{-\frac{\nu(R)}{2}}
  n_{b}(R)\right. \nonumber \\ & &
  +\left.[\xi(R_{s},\theta,t=0)-\xi(R_{s},\theta,t)]
  \int_{R_{s}-\Delta r}^{R_{s}}dR \frac{d}{dR}
  \left[R^{2}{\rm e}^{-\frac{\nu(R)}{2}}\right]n_{b}(R)
  \right\}\ . \nonumber \\ & &
\end{eqnarray}

We now turn to the evaluations of changes in the pressure 
of matter in the crust induced by the spin-down. Recall
the definition of $\eta(\Delta r,\theta,t)$ shown in Figure 5 and
the evolving pressure profile of the crust described in \S\ 3.2.
Then, the change until an age $t$ in the pressure of 
an element that is initially located at the point denoted by $\Delta r$ 
and $\theta$ may be written as
\begin{eqnarray}
  \delta P(\Delta r,\theta,t) &=& P(R)|_{R=R_{s}-\Delta r-
  \eta(\Delta r,\theta,t)}-P(R)|_{R=R_{s}-\Delta r}
  \nonumber \\ &=&
  -\eta(\Delta r,\theta,t)\left[\frac{dP(R)}{dR}\right]_{R=R_{s}-\Delta r}
  +{\rm O}(\Omega^{4})\ ,
\end{eqnarray}
where $P(R)$ is the pressure profile of the non-rotating 
configuration shown in Figure 3.
We have calculated the values of $\eta(\Delta r,\theta,t)$
according to equation (56), and, by substituting these values
into equation (57), we have obtained the values of 
$\delta P(\Delta r,\theta,t)$. 
The result has been expressed as
\begin{equation}
  \frac{\delta P(\Delta r,\theta,t)}{P(R_{s}-\Delta r)}
  =(-0.11+0.35\sin^{2}\theta)\frac{\Omega(t=0)^{2}-\Omega(t)^{2}}
  {\Omega_{0}^{2}}\ .
\end{equation}
Here we have omitted the $\Delta r$ dependence of 
$\delta P(\Delta r,\theta,t)/P(R_{s}-\Delta r)$, since it makes 
negligible differences. 

Figure 6 illustrates as a function of $\theta$ the values of
$\delta P(\Delta r,\theta,t\rightarrow\infty)/P(R_{s}-\Delta r)$
given by equation (58) in which we set $P_{\rm rot}(t=0)=1$ ms;
this quantity denotes the change in the pressure of 
an element of the crustal matter
at an age of far later than $\tau_{s}$ 
divided by its initial pressure.
We thus see
that the matter is compressed in the equatorial region of the crust and
is decompressed in the polar region due to the spin-down, as noted in
\S\ 1.

\vspace*{1.0cm}

\noindent
\begin{center}
{\bf 4.~ TRANSITIONS OF NUCLEAR COMPOSITIONS IN THE CRUST}
\end{center}
\vspace*{0.5cm}

Let us consider nuclear processes induced in the cold crustal 
matter, initially in nuclear equilibrium, by the changes in 
its pressure due to the spin-down as discussed in the preceding
section. We first ask how the matter being compressed or 
decompressed departs from its ground state in which the system 
satisfies the equilibrium conditions
(23)--(25), (27), and (29). We note that the time scale 
for pressure readjustment, corresponding to the sound travel 
time across a nucleus (Lattimer et al.\ 1977), is negligibly small 
compared with the spin-down time scale $\tau_{s}$, so that the system 
should always satisfy the pressure conditions (23) and (24).
Due to the assumption that the system contains a single species
of nucleus at a given pressure, 
we may replace the condition (25) by the condition 
that the number of baryons, $A_{\rm cell}$, in a single cell 
remains constant. If nuclear fusion (fission) were to occur,
however, this process would double (halve) the value of 
$A_{\rm cell}$. The conditions (27) and (29) ensure 
stability of the matter against $\beta-$ and neutron processes, 
respectively. While one of these processes is in progress, 
the system, not satisfying the corresponding stability condition,
is out of quasi-equilibrium.
Here we note that the time scale for neutron processes is
far shorter than $\tau_{s}$, and we assume that the time scale 
for $\beta-$processes is also smaller than $\tau_{s}$.
Then, the system may be considered to retain quasi-equilibrium 
during the spin-down. We shall thus pay 
attention only to the energy criterion for the occurrence of
nuclear processes in analyzing the chemical evolution of the crust.

By noting the Gibbs free energy of the system per nucleon 
given by equations (11) and (18) as well as the conditions
to be satisfied in the compression and decompression
as discussed above, 
we have examined transitions of the composition of a unit cell 
from the equilibrium one listed in Table 1
due to the change in its pressure during the spin-down as follows:
(i) At each pressure shifting from the equilibrium value $P_{\rm eq}$,
we have calculated the Gibbs free energy per nucleon under the
charge neutrality (22) and the pressure conditions (23) and (24) for
the initial nuclide $(A,Z,A_{\rm cell})$ as well as for
the nuclides $(A,Z\pm 1,A_{\rm cell})$, $(A\pm 1,Z,A_{\rm cell})$,
$(2A,2Z,2A_{\rm cell})$, and $(A/2,Z/2,A_{\rm cell}/2)$,
resulting from $\beta$-processes, neutron processes,
fusion, and fission, respectively. (ii) By finding 
the case in which the Gibbs free energy for the nuclide
arising from the reactions
becomes equal to the energy for the initial one,
we have determined the type of reaction expected 
to occur first and its threshold pressure $P_{\rm th}$. 
(iii) At the pressure equal to $P_{\rm th}$, we have determined 
the transitions to the next nuclides according to the criterion that 
the Gibbs free energy calculated for the nuclide present 
after the reactions under the conditions (22)--(24) 
should be lower than that for the previous one,
until the system attains quasi-equilibrium.
(iv) By following a line similar to (i)--(iii), we have examined
a series of processes expected to occur with a further change
in pressure. 
In (i)--(iv) we have assumed that, if $\beta$- and neutron processes 
were energetically possible at the same time, neutron 
processes would go first. 

In addition to the changes in the nuclear compositions, 
we have evaluated the nuclear energy $q$ thereby 
released per nucleon according to the thermodynamics at
zero temperature as $q=g_{i}-g_{f}$, 
where $g_{i}$ is 
the Gibbs free energy per nucleon for the nuclide
before the reactions, and $g_{f}$ is 
that for the next nuclide.
As far as $\beta-$ and neutron processes are concerned,
not only the reactions which release no energy but
those which release some can take place.
If the Fermi energies of electrons (dripped neutrons) become 
equal to the energy thresholds for $\beta-$ (neutron) processes
during the compression or decompression,
the electrons (neutrons) at the sea levels 
disappear or come out without supplying the medium with energy.    
Since these reactions proceed in a quasi-equilibrium condition,
we shall refer to such processes as quasi-equilibrium processes.
On the other hand, the reactions can liberate some energy
when the Fermi energies of electrons (dripped neutrons) 
are in excess of the energy thresholds
for electron captures (neutron absorption), i.e., $\mu_{e}>W_{Z-1}$
$(\mu_{n}+m_{n}c^{2}>D_{A+1})$,
as well as when those are below the energy 
thresholds for $\beta-$decays (neutron emission), i.e., 
$\mu_{e}<W_{Z}$ $(\mu_{n}+m_{n}c^{2}<D_{A})$.
In the former case, the electrons (dripped neutrons)
with energy between $W_{Z-1}$ $(D_{A+1})$ and 
$\mu_{e}$ $(\mu_{n}+m_{n}c^{2})$ are captured by the nuclei,
which leads to excitation of the nuclei and to creation of
holes in the electron (neutron) Fermi sphere.
In the latter case, 
the nuclei add the electrons (neutrons) with
energy between $\mu_{e}$ $(\mu_{n}+m_{n}c^{2})$ and $W_{Z}$ 
$(D_{A})$ to the electron (neutron) Fermi sea and simultaneously
undergo excitation.
In both cases, therefore, the reactions reduce the system
to a non-equilibrium state; hereafter we shall denote such
processes by non-equilibrium processes.  
At zero temperature considered here, the following relaxation towards 
quasi-equilibrium results in liberation of
the energy $q A_{\rm cell}$ per unit cell;
for neutron absorption, neutron emission,
electron captures, and $\beta-$decays, 
$q A_{\rm cell}\approx\mu_{n}+m_{n}c^{2}-D_{A+1}$,
$q A_{\rm cell}\approx D_{A}-\mu_{n}-m_{n}c^{2}$, 
$q A_{\rm cell}\approx\mu_{e}-W_{Z-1}$, and
$q A_{\rm cell}\approx W_{Z}-\mu_{e}$, respectively.
The relaxation processes involved, the resulting heat generation,
and the applicability of the zero-temperature approximation
will be discussed in \S\ 5.

We proceed to show the results for the transitions from the
equilibrium nuclides to the other ones
during the compression of matter in the equatorial region of
the crust due to the spin-down. The increase in its pressure 
can amount to about $24$\% 
for $P_{\rm rot}(t=0)=1$ ms as shown in Figure 6. In Figure 7
we have drawn the diagrams that show the changes in nuclides
induced by such or a little larger increase in
pressure from six equilibrium nuclides present at mass densities
$\rho_{\rm eq}$ above neutron drip. 

In the initial density region $4.3\times10^{11}$ g cm$^{-3}
\leq\rho_{\rm eq}\lsim 3\times10^{13}$ g cm$^{-3}$, 
as shown in Figures 7a--7c,
electron captures occur first 
when $P=P_{\rm th}$, at which the electron chemical potential 
$\mu_{e}$, increasing due to the compression, attains the
energy threshold $W_{Z-1}$. 
These processes are
quasi-equilibrium ones. Thereafter, a succession of neutron
emission and electron captures takes place
at the same pressure as non-equilibrium processes. 
This result indicates that the nuclei undergoing non-equilibrium
neutron emission (electron captures) have energy thresholds 
$D_{A}$ ($W_{Z-1}$) larger (smaller) than $\mu_{n}+m_{n}c^{2}$ 
$(\mu_{e})$ in contrast with the initial nuclei. This contrast is
partly because the initial nuclei have even proton and neutron
numbers and hence receive greater stability
from the pairing effects, which are described by equation (10), 
and partly because they have closed shells of protons
leading to local energy minima, which are determined by equation (6). 
A quasi-equilibrium state, in which the system 
includes the even-even nuclei in most cases, follows this series 
of reactions and lasts until the pressure
arrives at the threshold for the forthcoming electron captures.  

For initial densities of about $3\times10^{13}$
g cm$^{-3}$, neutron absorption begins to take place along with
$\beta$-processes as a result of the increase in the chemical 
potential of dripped neutrons during the compression,
as can be seen from Figure 7d.
This result implies that, at such initial densities,
the nuclei are immersed in the neutron gas dense enough for the 
threshold pressure for neutron absorption to become comparable 
to that for electron captures.
For higher initial densities up to about $2\times10^{14}$ g 
cm$^{-3}$, as shown in Figures 7e and 7f, neutron absorption rather 
than electron captures occurs first as quasi-equilibrium processes 
when $P=P_{\rm th}$ at which $\mu_{n}+m_{n}c^{2}\approx D_{A+1}$. 
These processes are accompanied by neutron absorption 
which occurs at the same pressure in a non-equilibrium 
condition and hence releases some nuclear energy. 
The resulting state in which the system contains
the even-even nuclei remains in quasi-equilibrium, until 
further increase in pressure leads to the next 
two-stage neutron absorption.
The two-step feature of the reactions stems from the fact that
the energy thresholds $D_{A+1}$ depend on the parity of 
the reacting nuclei due to the nuclear pairing effects;
this feature can be observed schematically from Figure 8a.

At initial densities below neutron drip,
it is shown that electron captures
take place first at the threshold pressure 
higher than $P_{\rm eq}$ by more than $40$\%. Such increment in
pressure, however, is larger than that expected from the 
compression due to the spin-down. 
This result
suggests that the energy thresholds for electron captures, $W_{Z-1}$,
formed mainly by the closed shells of neutrons (protons for Ni) 
and pairing gaps in the equilibrium nuclei present 
in this density regime (see Table 1) are large enough to keep
the nuclei stable against the reactions during the spin-down.

As can be observed from the results mentioned above,
no fusion reaction is allowed energetically 
by the compression during the spin-down 
in the whole density region of the crust. 
This is because electron captures, 
leading to a reduction in the nuclear charge, 
do not proceed sufficiently for the occurrence
of fusion.

We now summarize the nuclear processes expected to occur in 
the compressed matter during the spin-down and the energies 
thereby released.
By comparing the threshold pressure $P_{\rm th}$ for the 
processes occurring first (see Figure 7) with the pressure to
be increased by the spin-down with $P_{\rm rot}(t=0)=1$ ms in the
equatorial region of the crust (see Figure 6), we find
that, when $P_{\rm rot}(t=0) \gsim 1$ ms,  
almost all the processes occurring in the compression are 
neutron absorption by the nuclei initially present in the density region 
$3\times10^{13}$ g ${\rm cm}^{-3}\lsim\rho_{\rm eq}\lsim 2\times10^{14}$ g
${\rm cm}^{-3}$. In Table 2 we have shown the changes in
the nuclear compositions in such a density regime with
increasing pressure up to $24$\% and the energy $q$ released
per nucleon. 
We see that both the ratio of
$P_{\rm th}$ to $P_{\rm eq}$ shown in Figure 7 and the
energy released by non-equilibrium neutron absorption
as listed in Table 2 tend to decrease as the initial density 
(proton fraction)
increases (decreases), a feature coming primarily from
the isospin dependence of $W_{\rm pair}$
given by equation (10).
   
For the purpose of examining the change in the equation of 
state of the compressed matter from the equilibrium one 
during the spin-down, 
we have also calculated the equation of state 
for the compositions of matter which are altered 
from the equilibrium ones listed in Table 1
by an increase of $30$\% in pressure.
Figure 9 depicts the resulting equation of state,
together with the equilibrium equation of state listed in
Table 1 and with the equation of state of BPS\@. We observe 
in this figure that the non-equilibrium equation of state, 
as well as the equilibrium one, 
agrees graphically with
that of BPS in the nearly whole density region.
Such agreement between the non-equilibrium and
BPS equations of state may justify 
our neglecting effects of the variation in the
equation of state during the spin-down on
the structure of the crust, as noted in \S\ 3.2. 
It is interesting to notice
the deviation of the non-equilibrium equation of state
from the equilibrium one found for densities between
$4.3\times10^{11}$ and $3\times10^{12}$ g ${\rm cm}^{-3}$.
Such deviation originates from the fact that the threshold effects
prevent neutron emission, leading to softening of the 
equation of state in this density range, from proceeding
enough for the number density of dripped neutrons $n_{n}$ to
reach its equilibrium value at the same pressure,
as suggested by Haensel \& Zdunik (1990b). 

We conclude this section by showing the results for 
the changing compositions of the decompressed matter 
in the polar region of the crust during the spin-down. 
We have found that, basically, the processes occurring 
in the decompression are inverse processes against those 
caused by the compression in the whole density regime. 
This result is due to the fact that 
the electron and external neutron gases, controlling
the occurrence of $\beta$- and neutron processes,
undergo a reduction in their chemical potentials during 
the decompression in contrast with the case during the 
compression.
For $P_{\rm rot}(t=0)=1$ ms, Figure 6 shows that
the pressure of the matter in the vicinity of the rotation 
axis decreases by about $11$\% during the spin-down. 
In consequence of such decrease in pressure, neutron 
emission is found to take place as two-stage reactions
composed of a quasi-equilibrium process and a non-equilibrium 
one at initial densities
between $5\times10^{13}$ and $2\times10^{14}$ g ${\rm cm}^{-3}$.
Figure 8b depicts a schematic of nuclear levels relevant to
these two-stage reactions. 
In Table 3 we have shown the changes in the nuclear 
compositions in such a density regime with decreasing pressure
down to $11$\% and the energy per nucleon released by neutron
emission. In comparison with Table 2, we observe that
the changes in nuclides during the decompression and those 
during the compression from the same equilibrium nuclides
are analogous with each other
in terms of the energy release and of 
the pressure intervals between the adjacent thresholds for 
the reactions.

\vspace*{1.0cm}

\noindent
\begin{center}
{\bf 5.~ HEAT GENERATION AND NEUTRON STAR THERMAL EVOLUTION}
\end{center}
\vspace*{0.5cm}

The energies released by neutron absorption and emission that
occur during the compression and decompression turn
completely into heat through relaxation processes
such as scattering between neutrons in the continuum states
and $\gamma$ de-excitation of the reacting nuclei.
The heat thereby deposited locally in the substance
is $\sim 0.1$ MeV per reaction, which is in turn
carried away by electrons through their
scattering with phonons, impurities, and other electrons 
(Flowers \& Itoh 1976). 
The time scale for the heat conduction across a unit cell can
be estimated as $\tau_{H}\sim c_{P}r_{c}^{2}/\kappa$,
where $c_{P}$ is the specific heat per unit volume at constant
pressure, which is simply assumed to be
the sum of that for the
degenerate nonrelativistic neutrons of density $n_{n}$ and that
for the degenerate relativistic electrons of density $n_{e}$,
and $\kappa$ is the thermal conductivity.
We then note that the time interval, $\tau_{I}$, between when
the pressure reaches the threshold for the reaction in a cell and
when in the neighboring cell is $\sim \tau_{s}r_{c}/\eta$, where
$\eta$ is given by equation (56).
At typical densities and temperatures of the substance 
around the reacting cell, 
$c_{P}\sim 10^{20}$ erg K$^{-1}$
cm$^{-3}$, $\kappa\sim 10^{20}$ erg cm$^{-1}$ K$^{-1}$ s$^{-1}$,
$r_{c}\sim 10$ fm, and $\eta\sim 10$ m, so that $\tau_{H}\ll\tau_{I}$. 
We can thus assume that the heat production by the reaction 
in a cell makes no change in the energy criterion 
for the reactions occurring next in the surrounding cells, which 
has been designed at zero temperature in \S\ 4, and hence in the
results shown in Tables 2 and 3 
for the chemical evolution of the crust. 

The total heat, $Q_{\rm tot}$,
deposited in the star during the spin-down with $P_{\rm rot}(t=0)=1$ ms
may be estimated from the following: (i) The energy
per unit volume released by a non-equilibrium neutron process
that occurs in the matter initially present at densities between 
$3\times10^{13}$ and $2\times10^{14}$ g ${\rm cm}^{-3}$
$(0.02$ fm$^{-3} \lsim n_{b}\lsim 0.1$ fm$^{-3})$
is evaluated from Tables 2 and 3 as $n_{b}q\approx 4$ eV fm$^{-3}$,
irrespective of densities.
(ii) The reacting material is initially located at the
interior of depths $\Delta r$ 
between $450$ and $700$ m,
as can be seen from Figure 3. (iii) The number of the non-equilibrium 
processes, $N(\Delta r)$, occurring during the spin-down 
in a unit cell that is initially located at the depth $\Delta r$
on the equatorial plane is estimated from Table 2 and Figure 3 as 
$N(\Delta r)\approx 15(\Delta r-r_{1})/(r_{2}-r_{1})$, 
$r_{1}<\Delta r<r_{2}$, where $r_{1}=450$ m and $r_{2}=700$ m.  
(iv) 
The ratio between the number of 
the non-equilibrium processes taking place on the 
cone of fixed $\theta$ and that on the equatorial plane is
$\approx \sin\theta|1-1.46\cos^{2}\theta|$
as deduced from equation (58). 
By using (i)--(iv) and by taking account of the redshift effect
on energy generation inside the non-rotating star (Thorn 1977),
we have roughly estimated $Q_{\rm tot}$ as
\begin{eqnarray}
  Q_{\rm tot} &\approx& 
  \frac{2\pi R_{s}^{2}n_{b}q}
  {\sqrt{1-\frac{2GM}{R_{s}c^{2}}}}
  \int_{r_{1}}^{r_{2}}d\Delta r N(\Delta r)
  \int_{0}^{\pi}d\theta\sin\theta|1-1.46\cos^{2}\theta|
  \nonumber \\ 
  &\approx& 
  10^{46}\ {\rm erg}\ .
\end{eqnarray}
This value of $Q_{\rm tot}$ is found to be smaller than
the initial rotational energy of the star, $I\Omega_{0}^{2}/2$, 
by six orders of magnitude;
therefore, $Q_{\rm tot}$ has negligible effects on the rotational 
evolution determined by the magnetic dipole braking.

We now consider a heating rate $H(t)$, the heat deposited
in the star by the non-equilibrium neutron processes 
that occur per unit time at an age $t$.
Note that both the pressure intervals 
between the adjacent thresholds for the non-equilibrium reactions
occurring in a unit cell
and the energies thereby released 
take on almost constant values,
as can be seen from Tables 2 and 3.
Then, we may assume that
the heat generated until a given age is proportional
to the variation in the pressure of the reacting material 
due to the spin-down until that age 
as described by equation (58). We thus obtain
\begin{equation}
  \int_{0}^{t}H(t')dt'=\frac{Q_{\rm tot}[\Omega(t=0)^{2}-\Omega(t)^{2}]}
  {\Omega_{0}^{2}}\ .
\end{equation}
Consequently, the heating rate may be expressed as 
\begin{eqnarray}
  H(t) &\approx& 5\times10^{-7}{\dot E}(t) \nonumber \\ 
  &\approx& 5\times10^{32}\frac{\left[\frac{B({\rm G})}
  {10^{10}{\rm G}}\right]^{2}\frac{\Omega(t=0)^{4}}{\Omega_{0}^{4}}}
  {\left\{1+1.54\left[\frac{B({\rm G})}{10^{10}{\rm G}}\right]^{2}
  \frac{\Omega(t=0)^{2}}{\Omega_{0}^{2}}
  \frac{t({\rm yr})}{10^{6}{\rm yr}}\right\}^{2}}
  \ {\rm erg}\ {\rm s}^{-1}\ ,
\end{eqnarray}
where   
\begin{equation}
  {\dot E}(t)=-I\Omega(t){\dot \Omega}(t)
\end{equation}
is the spin-down power derived from equation (51).

To ascertain the practical utility of equation (61), we recall that
the occurrence of neutron processes in the crustal matter requires 
the changes of at least a few percent in its pressure, 
as seen from Tables 2 and 3.
In consequence, the condition for use of equation (61)
is roughly estimated from equation (58) as
\begin{equation}
  \Omega(t=0)^{2}-\Omega(t)^{2}\gsim 0.1 {\Omega_{0}^{2}}\ .
\end{equation}
We thus see that the present heating mechanism works when
$P_{\rm rot}(t=0)\lsim 3$ ms and $t\gsim 0.1 \tau_{s}$.

Figure 10 illustrates as a function of the age
the heating rate given by equation (61) in which $P_{\rm rot}(t=0)=1$ ms
and $B=10^{8}, 10^{9}, 10^{10}, 10^{11}, 10^{12}$ G\@. 
The age at which $H(t)$ starts to decline remarkably 
corresponds to the spin-down time scale $\tau_{s}$. 
We first consider the age of up to $\sim 10^{6}$ yr when
neutrino emission from the interior of the star dominates 
over photon emission from the surface 
(see, e.g., Nomoto \& Tsuruta 1987). 
In the light of models for the neutron star thermal evolution 
including some heating sources (see, e.g., Cheng et al.\ 1992; 
Reisenegger 1995; Van Riper, Link, \& Epstein 1995),
we expect that, at such times, 
the present heating source
inside the stars even with high magnetic fields $\sim 10^{12}$ G
does not contribute to the surface emission considerably 
compared with the contribution of the initial cooling of the stars. 
For $t>10^{6}$ yr, on the other hand,
it is generally predicted that the heat deposited initially is
gone, and that the surface emission, dominant over the
neutrino emission, is controlled entirely by some heat 
generation inside the star. 
It is noteworthy in Figure 10 that the heating rate for 
$B\sim 10^{8}$ G lasts during a term far longer than $10^{6}$ yr. 
We thus see that the old neutron stars with low magnetic fields, 
characteristic of millisecond pulsars, are the most significant 
examples in investigating effects of the present heating mechanism 
on the neutron star thermal evolution.

In Figure 11 we have depicted the bolometric and X-ray (0.1--2.4 keV)
luminosities, $L_{\rm bol}^{H}$ and $L_{X}^{H}$, predicted from
the heating rate (61)
as a function of the spin-down power $\dot{E}$, together with
the X-ray luminosities obtained from the {\it ROSAT} observations
of some millisecond pulsars in the energy range $0.1-2.4$ keV
(Danner, Kulkarni, \& Thorsett 1994) and with an empirical formula
for X-ray luminosities coming from the stellar magnetosphere
as proposed by \"{O}gelman (1995). Here we have evaluated the
bolometric luminosity $L_{\rm bol}^{H}$ by assuming a complete balance
between the photon emission from the surface of the star and the heat
production inside the star as well as by taking into account the
redshift effect on the surface emission
in the configuration of the non-rotating star
(Thorn 1977) as
\begin{equation}
  L_{\rm bol}^{H}=H\left(1-\frac{2GM}{R_{s}c^{2}}\right)\ .
\end{equation}
The X-ray luminosity $L_{X}^{H}$ has been then obtained
by assuming a spectrum of the emission from a spherical
blackbody of radius $R_{s}$ having the uniform
surface temperature as observed by a distant spectator
(Nomoto \& Tsuruta 1987):
\begin{equation}
  T_{s}^{\infty}=\left(\frac{L_{\rm bol}^{H}}{4\pi\sigma_{\rm SB}
  R_{s}^{2}}\right)^{1/4}
  \left(1-\frac{2GM}{R_{s}c^{2}}\right)^{1/4}\ ,
\end{equation}
where $\sigma_{\rm SB}$ is the Stefan-Boltzmann constant.  
We have selected the pulsars satisfying the condition (63) among
those chosen by Danner et al.\ (1994).

We observe in Figure 11 that the X-ray luminosities of millisecond
pulsars are far larger than $L_{X}^{H}$.
In comparison with the empirical result for
magnetospheric X-ray luminosities obtained by \"{O}gelman (1995),
however, the X-ray luminosities observed are predicted to
contain a large amount of nonthermal magnetospheric component
in addition to a thermal component. 
We may thus conclude that the heat 
deposited in the star by neutron absorption and emission during
the spin-down contributes little to the X-ray luminosities from
millisecond pulsars; such a contribution is
buried in the other components such as the magnetospheric one.
  
\vspace*{1.0cm}

\noindent
\begin{center}
{\bf 6.~ CONCLUSIONS}
\end{center}
\vspace*{0.5cm}

We have calculated the changes in the pressure of zero-temperature
matter in the crust of an isolated rotating
neutron star during its spin-down and the resulting transitions of
the compositions of the matter 
initially in nuclear equilibrium. 
We have found that the matter is compressed and
decompressed in the equatorial and polar regions of the crust, 
respectively, as the rotation of the star is slowing down.
The corresponding change in the equation of state has
been shown to have negligible effects on the structure of the crust.
Through the compression (decompression),
the nuclei initially present at densities between
$3\times 10^{13}$ and $2\times 10^{14}$ g cm$^{-3}$
have been found to absorb (emit) neutrons, when 
$P_{\rm rot}(t=0)\lsim 3$ ms.
The heating rate arising from these processes has been
estimated; we have clarified its marked effects on the thermal 
evolution of old neutron stars with low magnetic fields 
of $\sim 10^{8}$ G\@. 
It has been shown that the X-ray (0.1--2.4 keV)
luminosity predicted from
the present heating mechanism is far smaller than the values
observed from millisecond pulsars.
   
The uncertainty affecting the occurrence of nuclear processes,
which has been found in the dripped-neutron regime, stems
primarily from the present extrapolation of the empirical formula
for nuclear pairing gaps to higher relative neutron excess $\delta$,
as noted in \S\ 2, and possibly from the omission of pairing
correlations in the external neutron gas. Fortunately, the pairing
effects in the neutron gas, leading to stabilization of the continuum
states of neutrons, would be insensitive to the energy thresholds 
for reactions, 
to which, however, the pairing correlations between nucleons inside
a nucleus are crucial. We expect that the validity of
the extrapolation used here will be confirmed
by microscopic calculations of the pairing gaps
in the nuclei far from the $\beta$-stability valley with a 
realistic nucleon-nucleon interaction.

Another problem which remains to be investigated is the detailed
description of the transition from roughly spherical nuclei 
in the external neutron gas to uniform nuclear matter. 
Lorenz et al.\ (1993) and Oyamatsu (1993) discovered that
the phases containing nuclei of nonspherical (cylindrical or planar) 
shapes may exist at densities lower than the point 
at which the matter becomes uniform. 
If so, these nuclear shapes, affecting the discrete states of 
nucleons inside a nucleus, may have nonnegligible consequences 
for the occurrence of nuclear processes and hence
for the heat generation.
\vspace*{1.5cm}

We are grateful to Dr. T. Takatsuka and Dr. Y. Furihata for useful
discussion and comments. This work was supported in part by
Grants-in-Aid for Scientific Research provided by the Ministry of
Education, Science, and Culture of Japan through Research 
Grants Nos. 05243103 and 07CE2002.

\vspace*{1.0cm}

\newpage

\begin{table}[h]
\begin{center}
\begin{minipage}[t]{10cm}
\begin{center}
TABLE 1
\end{center}
\begin{center}
P{\scriptsize ROPERTIES} {\scriptsize OF} 
{\scriptsize THE} G{\scriptsize ROUND}-S{\scriptsize TATE}
M{\scriptsize ATTER} {\scriptsize IN} {\scriptsize THE}
C{\scriptsize RUST}
\end{center}
\end{minipage}
\end{center}
\begin{center}
\begin{tabular}{ccccccccc} \hline\hline
  $P$  & $\rho$  
  & $n_{b}$  &  &  &  &  &
  $\mu_{e}$  & $\mu_{n}$ \\
(MeV fm$^{-3}$) & (g cm$^{-3}$)& (fm$^{-3}$) & $Z$ & $A$ &
$A_{\rm{cell}}$ & $x$ & (MeV) & (MeV)\\ \hline
 & & $outer$ & & $crust$ & & & &\\
\hline
  4.000$\times 10^{-7}$ & 1.496$\times 10^{9}$ & 9.005$\times 10^{-7}$
  & 28 & 66 & 66. &0.42& 4.459 & 0 \\
  1.000$\times 10^{-6}$ & 3.022$\times 10^{9}$ & 1.819$\times 10^{-6}$
  & 36 & 86 & 86. &0.42& 5.598 & 0 \\
  5.000$\times 10^{-6}$ & 1.041$\times 10^{10}$ & 6.259$\times 10^{-6}$
  & 34 & 84 & 84. &0.40& 8.338 & 0 \\
  1.000$\times 10^{-5}$ & 1.812$\times 10^{10}$ & 1.089$\times 10^{-5}$
  & 32 & 82 & 82. &0.39& 9.902 & 0 \\
  3.000$\times 10^{-5}$ & 4.293$\times 10^{10}$ & 2.578$\times 10^{-5}$
  & 30 & 80 & 80. &0.38& 13.01 & 0 \\
  8.000$\times 10^{-5}$ & 9.349$\times 10^{10}$ & 5.608$\times 10^{-5}$
  & 28 & 78 & 78. &0.36& 16.61 & 0 \\
  1.000$\times 10^{-4}$ & 1.150$\times 10^{11}$ & 6.897$\times 10^{-5}$
  & 44 & 126 & 126. &0.35& 17.63 & 0 \\
  1.500$\times 10^{-4}$ & 1.605$\times 10^{11}$ & 9.874$\times 10^{-5}$
  & 42 & 124 & 124. &0.34& 19.50 & 0 \\
  2.000$\times 10^{-4}$ & 2.055$\times 10^{11}$ & 1.231$\times 10^{-4}$
  & 40 & 122 & 122. &0.33& 20.95 & 0 \\
  3.000$\times 10^{-4}$ & 2.882$\times 10^{11}$ & 1.725$\times 10^{-4}$
  & 38 & 120 & 120. &0.32& 23.17 & 0 \\
  4.300$\times 10^{-4}$ & 3.914$\times 10^{11}$ & 2.342$\times 10^{-4}$
  & 36 & 118 & 118. &0.31& 25.34 & 0 \\
\hline
 & & $inner$ & & $crust$ & & & &\\
\hline                                                                      
  5.012$\times 10^{-4}$ & 4.667$\times 10^{11}$ & 2.791$\times 10^{-4}$
  & 40 & 130 & 139.7 &0.31& 26.29 & 0.124 \\
  5.635$\times 10^{-4}$ & 6.692$\times 10^{11}$ & 3.999$\times 10^{-4}$
  & 40 & 130 & 189.5 &0.31& 26.79 & 0.388 \\
  6.828$\times 10^{-4}$ & 1.004$\times 10^{12}$ & 6.000$\times 10^{-4}$
  & 40 & 132 & 263.1 &0.30& 27.48 & 0.648 \\
  8.746$\times 10^{-4}$ & 1.472$\times 10^{12}$ & 8.791$\times 10^{-4}$
  & 40 & 134 & 354.7 &0.30& 28.26 & 0.919 \\
  1.483$\times 10^{-3}$ & 2.663$\times 10^{12}$ & 1.589$\times 10^{-3}$
  & 40 & 136 & 547.9 &0.29& 29.78 & 1.434 \\
  4.055$\times 10^{-3}$ & 6.257$\times 10^{12}$ & 3.730$\times 10^{-3}$
  & 50 & 180 & 1156. &0.28& 33.23 & 2.461 \\
  7.346$\times 10^{-3}$ & 9.684$\times 10^{12}$ & 5.770$\times 10^{-3}$
  & 50 & 186 & 1412. &0.27& 35.96 & 3.179 \\
  1.371$\times 10^{-2}$ & 1.497$\times 10^{13}$ & 8.913$\times 10^{-3}$
  & 50 & 194 & 1627. &0.26& 39.65 & 4.069 \\
  4.746$\times 10^{-2}$ & 3.432$\times 10^{13}$ & 2.040$\times 10^{-2}$
  & 50 & 218 & 1834. &0.23& 50.20 & 6.448 \\
  0.1777 & 8.010$\times 10^{13}$ & 4.749$\times 10^{-2}$
  & 40 & 220 & 1417. &0.18& 67.32 & 10.38 \\
  0.4094 & 1.334$\times 10^{14}$ & 7.890$\times 10^{-2}$
  & 40 & 282 & 1300. &0.14& 82.06 & 14.09 \\
\hline
\end{tabular}
\end{center} 
\end{table}  

\begin{table}
\begin{center}
\begin{minipage}[t]{10cm}
\vspace*{-1.7cm}
\begin{center}
TABLE 2
\end{center}
\begin{center}
N{\scriptsize UCLEAR} C{\scriptsize OMPOSITIONS} {\scriptsize IN}
{\scriptsize THE} I{\scriptsize NNER} C{\scriptsize RUST}
{\scriptsize DURING} {\scriptsize THE} C{\scriptsize OMPRESSION}
\end{center}
\end{minipage}
\end{center}
\begin{center}
\begin{tabular}{ccccccccc} \hline\hline
  $P$  & & $\rho$ 
  & & & & $\mu_{e}$ & $\mu_{n}$ & $q$ \\ 
  (MeV fm$^{-3}$) & $P/P_{\rm eq}$ &  (g $\rm{cm}^{-3}$) & 
  $Z$ & $A$ & $A_{\rm{cell}}$ & (MeV) & (MeV) & (keV)\\ \hline 
0.04746$^{\rm a}$&1& 3.432$\times 10^{13}$ & 50 & 218 & 1834. & 50.07 
& 6.448 & \\
0.05410&1.14& 3.739$\times 10^{13}$ & 49 & 218 & 1834. & 51.31 & 6.767
& 0.000 \\
0.05700&1.20& 3.867$\times 10^{13}$ & 49 & 219 & 1834. & 51.89 & 6.896
& 0.000 \\
0.05700&1.20& 3.869$\times 10^{13}$ & 49 & 220 & 1834. & 51.90 & 6.896
& 0.000 \\
0.05700&1.20& 3.864$\times 10^{13}$ & 50 & 220 & 1834. & 52.22 & 6.889
& 0.043 \\
0.05766&1.21& 3.896$\times 10^{13}$ & 49 & 220 & 1834. & 52.02 & 6.923
& 0.000 \\
\hline                                                                      
0.1777$^{\rm a}$ &1& 8.010$\times 10^{13}$ & 40 & 220 & 1417. & 67.32
& 10.38 & \\
0.1881 &1.06& 8.293$\times 10^{13}$ & 40 & 221 & 1417. & 68.10 & 10.60
& 0.000 \\
0.1881 &1.06& 8.293$\times 10^{13}$ & 40 & 222 & 1417. & 68.10 & 10.60
& 0.096 \\
0.1947 &1.10& 8.473$\times 10^{13}$ & 40 & 223 & 1417. & 68.59 & 10.74
& 0.000 \\
0.1947 &1.10& 8.473$\times 10^{13}$ & 40 & 224 & 1417. & 68.59 & 10.73
& 0.093 \\
0.2014 &1.13& 8.654$\times 10^{13}$ & 40 & 225 & 1417. & 69.07 & 10.87
& 0.000 \\
0.2014 &1.13& 8.654$\times 10^{13}$ & 40 & 226 & 1417. & 69.07 & 10.87
& 0.090 \\
0.2082 &1.17& 8.833$\times 10^{13}$ & 40 & 227 & 1417. & 69.54 & 11.01
& 0.000 \\
0.2082 &1.17& 8.833$\times 10^{13}$ & 40 & 228 & 1417. & 69.54 & 11.00
& 0.088 \\
0.2151 &1.21& 9.012$\times 10^{13}$ & 40 & 229 & 1417. & 70.01 & 11.14
& 0.000 \\
0.2151 &1.21& 9.015$\times 10^{13}$ & 40 & 230 & 1417. & 70.01 & 11.14
& 0.086 \\
\hline                                                                      
0.4094$^{\rm a}$ &1& 1.334$\times 10^{14}$ & 40 & 282 & 1300. & 82.06
& 14.09 & \\
0.4232 &1.03& 1.360$\times 10^{14}$ & 40 & 283 & 1300. & 82.58 & 14.26
& 0.000 \\
0.4232 &1.03& 1.360$\times 10^{14}$ & 40 & 284 & 1300. & 82.58 & 14.26
& 0.050 \\
0.4315 &1.05& 1.375$\times 10^{14}$ & 40 & 285 & 1300. & 82.89 & 14.37
& 0.000 \\
0.4315 &1.05& 1.375$\times 10^{14}$ & 40 & 286 & 1300. & 82.89 & 14.37
& 0.049 \\
0.4396 &1.07& 1.390$\times 10^{14}$ & 40 & 287 & 1300. & 83.19 & 14.47
& 0.000 \\
0.4396 &1.07& 1.390$\times 10^{14}$ & 40 & 288 & 1300. & 83.19 & 14.47
& 0.049 \\
0.4477 &1.09& 1.405$\times 10^{14}$ & 40 & 289 & 1300. & 83.48 & 14.57
& 0.000 \\
0.4477 &1.09& 1.405$\times 10^{14}$ & 40 & 290 & 1300. & 83.48 & 14.57
& 0.048 \\
0.4560 &1.11& 1.420$\times 10^{14}$ & 40 & 291 & 1300. & 83.78 & 14.67
& 0.000 \\
0.4560 &1.11& 1.421$\times 10^{14}$ & 40 & 292 & 1300. & 83.79 & 14.67
& 0.048 \\
0.4641 &1.13& 1.435$\times 10^{14}$ & 40 & 293 & 1300. & 84.07 & 14.77
& 0.000 \\
0.4641 &1.13& 1.435$\times 10^{14}$ & 40 & 294 & 1300. & 84.07 & 14.77
& 0.046 \\
0.4720 &1.15& 1.449$\times 10^{14}$ & 40 & 295 & 1300. & 84.35 & 14.86
& 0.000 \\
0.4720 &1.15& 1.450$\times 10^{14}$ & 40 & 296 & 1300. & 84.36 & 14.86
& 0.046 \\
0.4801 &1.17& 1.464$\times 10^{14}$ & 40 & 297 & 1300. & 84.63 & 14.96
& 0.000 \\
0.4801 &1.17& 1.464$\times 10^{14}$ & 40 & 298 & 1300. & 84.63 & 14.96
& 0.044 \\
0.4878 &1.19& 1.478$\times 10^{14}$ & 40 & 299 & 1300. & 84.89 & 15.05
& 0.000 \\
0.4878 &1.19& 1.478$\times 10^{14}$ & 40 & 300 & 1300. & 84.89 & 15.05
& 0.043 \\
0.4962 &1.21& 1.493$\times 10^{14}$ & 40 & 301 & 1300. & 85.18 & 15.15
& 0.000 \\
0.4962 &1.21& 1.493$\times 10^{14}$ & 40 & 302 & 1300. & 85.18 & 15.15
& 0.043 \\
0.5042 &1.23& 1.507$\times 10^{14}$ & 40 & 303 & 1300. & 85.43 & 15.24
& 0.000 \\
0.5042 &1.23& 1.507$\times 10^{14}$ & 40 & 304 & 1300. & 85.44 & 15.24
& 0.042 \\
\hline
\end{tabular}

\begin{flushleft}
\begin{minipage}[h]{6cm}
{$^{\rm a}$ Initial equilibrium nuclides. }
\end{minipage}
\end{flushleft}

\end{center}
\end{table}  

\begin{table}
\begin{center}
\begin{minipage}[t]{10cm}
\begin{center}
TABLE 3
\end{center}
\begin{center}
N{\scriptsize UCLEAR} C{\scriptsize OMPOSITIONS} {\scriptsize IN}
{\scriptsize THE} I{\scriptsize NNER} C{\scriptsize RUST}
{\scriptsize DURING} {\scriptsize THE} D{\scriptsize ECOMPRESSION}
\end{center}
\end{minipage}
\end{center}
\begin{center}
\begin{tabular}{ccccccccc} \hline\hline
  $P$ & & $\rho$ & & & &
  $\mu_{\rm{e}}$ & $\mu_{\rm{n}}$ & $q$ \\ 
  (MeV fm$^{-3}$) & $P/P_{\rm eq}$ & (g cm$^{-3}$) 
  & $Z$ & $A$ & $A_{\rm{cell}}$ & (MeV) & (MeV) & (keV)\\ \hline 
0.1777$^{\rm a}$&1& 8.010$\times10^{13}$ & 40 & 220 & 1417. & 67.32 &
10.38 & \\
0.1662&0.94& 7.688$\times10^{13}$ & 40 & 219 & 1417. & 66.41 & 10.13 &
0.000 \\
0.1662&0.94& 7.684$\times10^{13}$ & 40 & 218 & 1417. & 66.40 & 10.13 &
0.100 \\
0.1598&0.90& 7.499$\times10^{13}$ & 40 & 217 & 1417. & 65.86 & 9.985 &
0.000 \\
0.1598&0.90& 7.497$\times10^{13}$ & 40 & 216 & 1417. & 65.86 & 9.985 &
0.103 \\
\hline                                                                      
0.4094$^{\rm a}$&1& 1.334$\times10^{14}$ & 40 & 282 & 1300. & 82.06 &
14.09 & \\
0.3941&0.96& 1.305$\times10^{14}$ & 40 & 281 & 1300. & 81.46 & 13.88 &
0.000 \\
0.3941&0.96& 1.304$\times10^{14}$ & 40 & 280 & 1300. & 81.44 & 13.88 &
0.051 \\
0.3862&0.94& 1.289$\times10^{14}$ & 40 & 279 & 1300. & 81.13 & 13.78 &
0.000 \\
0.3862&0.94& 1.289$\times10^{14}$ & 40 & 278 & 1300. & 81.13 & 13.78 &
0.053 \\
0.3784&0.92& 1.274$\times10^{14}$ & 40 & 277 & 1300. & 80.81 & 13.67 &
0.000 \\
0.3784&0.92& 1.273$\times10^{14}$ & 40 & 276 & 1300. & 80.80 & 13.67 &
0.054 \\
0.3705&0.90& 1.258$\times10^{14}$ & 40 & 275 & 1300. & 80.47 & 13.56 &
0.000 \\
0.3705&0.90& 1.257$\times10^{14}$ & 40 & 274 & 1300. & 80.46 & 13.56 &
0.055 \\
\hline
\end{tabular}

\begin{flushleft}
\begin{minipage}[h]{6cm}
{$^{\rm a}$ Initial equilibrium nuclides. }
\end{minipage}
\end{flushleft}

\end{center} 
\end{table}  

\newpage
\noindent
\begin{center}
{\bf REFERENCES}
\end{center}
\vspace*{0.5cm}

\noindent
Audi, G., \& Wapstra, A. H. 1993, Nucl. Phys., A565, 1

\noindent
Baym, G., Bethe, H. A., \& Pethick, C. J. 1971, Nucl. Phys., 
A175, 225

\noindent
Baym, G., Pethick, C. J., \& Sutherland, P. 1971, ApJ, 
170, 299

\noindent
Bisnovatyi-Kogan, G. S., \& Seidov, Z. F. 1970, 
Soviet Astron., 14, 113

\noindent
Cheng, K. S., Chau, W. Y., Zhang, J. L., \& Chau, H. F. 1992, ApJ,
396, 135 

\noindent
Danner, R., Kulkarni, S. R., \& Thorsett, S. E. 1994,
ApJ, 436, L153

\noindent
Flowers, E., \& Itoh, N. 1976, ApJ, 206, 218

\noindent
Haensel, P., \& Pichon, B. 1994, A\&A, 283, 313 

\noindent
Haensel, P., \& Zdunik, J. L. 1990a, A\&A, 227, 431

\noindent
--------- . 1990b, A\&A, 229, 117

\noindent
Hartle, J. B. 1967, ApJ, 150, 1005

\noindent
--------- . 1970, ApJ, 161, 111

\noindent
Hartle, J. B., \& Thorn, K. S. 1968, ApJ, 153, 807

\noindent
Lattimer, J. M., Mackie, F., Ravenhall, D. G., \& Schramm, D. N.
1977, ApJ, 213, 225

\noindent
Lorenz, C. P., Revenhall, D. G., \& Pethick, C. J. 1993,
Phys. Rev. Lett., 70, 379

\noindent
Misner, C. W., \& Sharp, D. 1964, Phys. Rev., 136, B571

\noindent
M{\" o}ller, P., \& Nix, J. R. 1992, Nucl. Phys., A536, 20 

\noindent
M{\" o}ller, P., Nix, J. R., Myers, W. D., \& Swiatecki, W. J. 1995,
At. Data Nucl. Data \\ \hspace*{1cm}Tables, 59, 185

\noindent
Myers, W. D., \& Swiatecki, W. J. 1966,
Nucl. Phys., 81, 1

\noindent
Negele, J. W., \& Vautherin, D. 1973, Nucl. Phys., A207, 298 
 
\noindent
Nomoto, K., \& Tsuruta, S. 1987, ApJ, 312, 711

\noindent
{\" O}gelman, H. 1995, in The Lives of the Neutron Stars, ed.
M. A. Alpar, {\" U}. Kizilo{\u g}lu, 
\\ \hspace*{1cm}\& J. van Paradijs
(Kluwer: Dordrecht), 101

\noindent
Oyamatsu, K. 1993, Nucl. Phys., A561, 431

\noindent
Pethick, C. J., \& Ravenhall, D. G. 1995,
Annu. Rev. Nucl. Part. Sci., 45, 429
  
\noindent
Ravenhall, D. G., Bennett, C. D., \& Pethick, C. J. 1972, 
Phys. Rev. Lett., 28, 978

\noindent
Reisenegger, A. 1995, ApJ, 442, 749

\noindent
Sato, K. 1979, Prog. Theor. Phys., 62, 957

\noindent
Shapiro, S. L., \& Teukolsky, S. A. 1983,
Black Holes, White Dwarfs, and Neutron Stars 
\\ \hspace*{1cm}(New York: Wiley)

\noindent
Taylor J. H., Manchester, R. N., \& Lyne, A. G. 1993, ApJS, 
88, 529

\noindent
Thorn, K. S. 1977, ApJ, 212, 825

\noindent
Van Riper, K. A., Link, B., \& Epstein, R. I. 1995, ApJ, 448, 294

\newpage

\begin{center}
{\bf FIGURE CAPTIONS}
\end{center}
\vspace*{0.5cm}

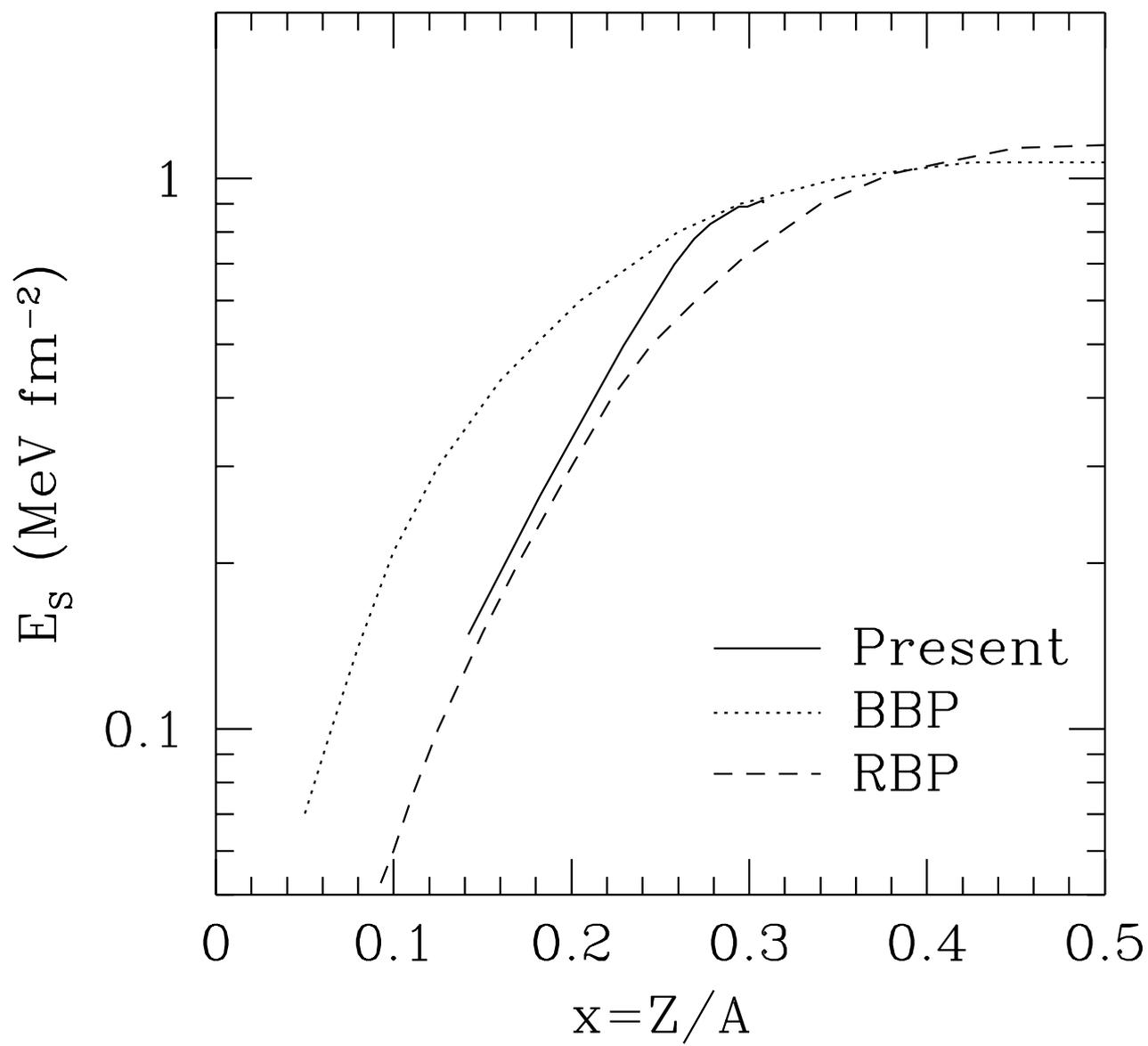
FIG. 1.-----  The surface tension as a function of $x$, the proton
fraction in nuclei. The solid curve is the result obtained in
the present calculations, the dotted curve is the BBP result, 
and the dashed curve is the RBP result.

\vspace*{0.5cm}

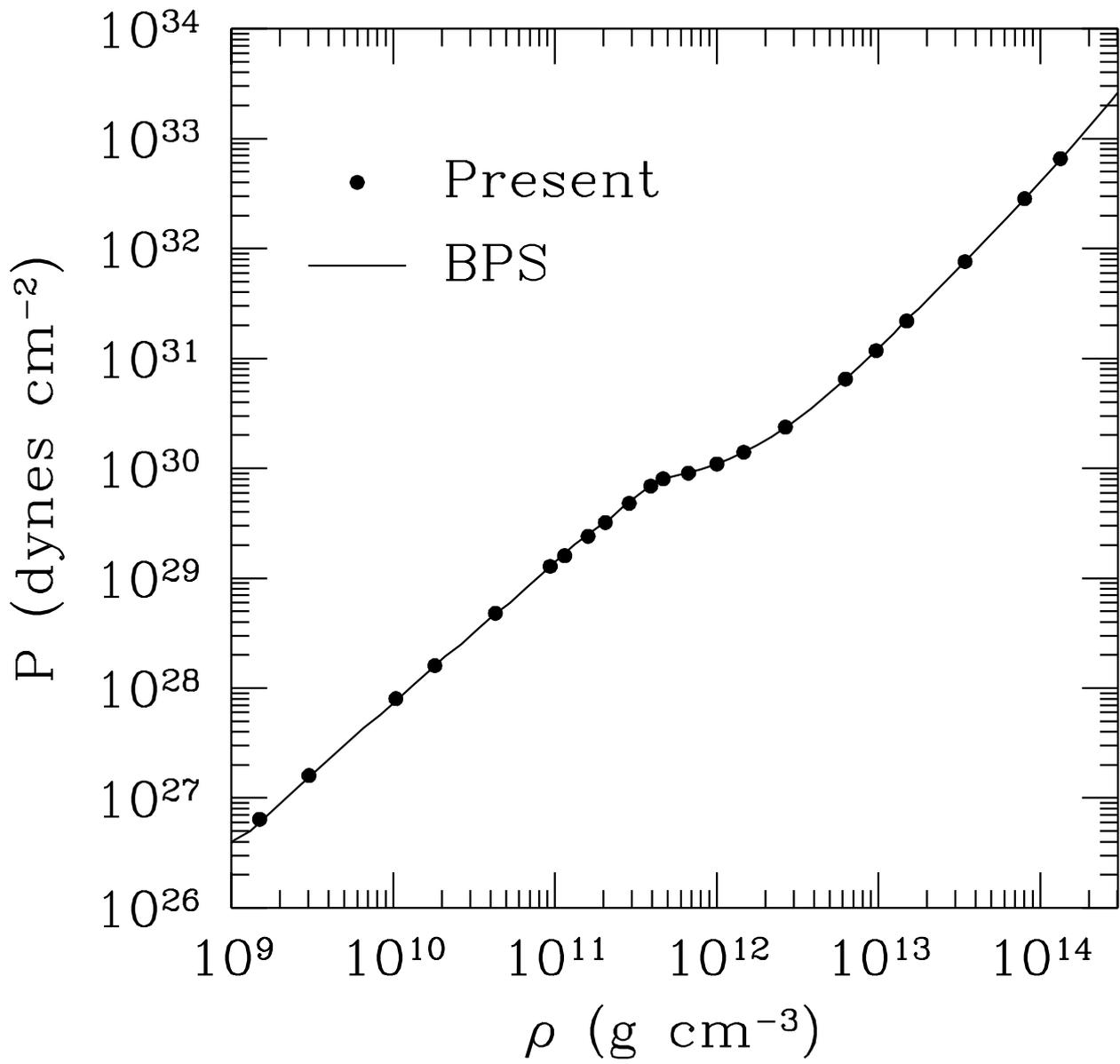
FIG. 2.-----  Equation of state of the crustal matter in its ground
state. The solid circles denote the present result, and the line 
is the BPS result.
 
\vspace*{0.5cm}

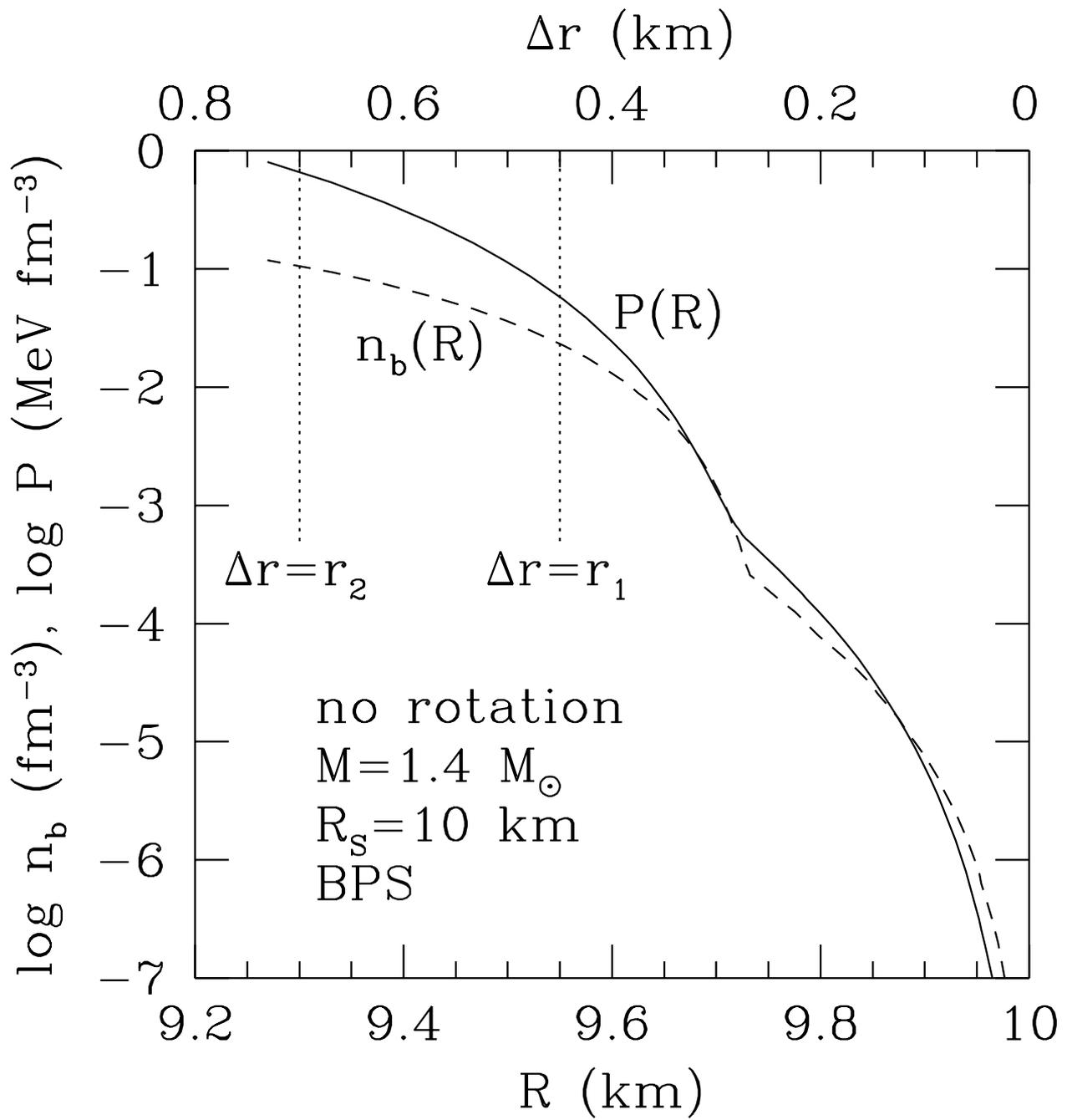
FIG. 3.-----  Pressure and density profiles of the crust of a
non-rotating neutron star of mass 1.4 $M_{\odot}$ and radius 10 km,
as calculated from the equation of state of BPS\@. The solid line is
the pressure $P$ (in MeV fm$^{-3}$), and the dashed line is the
number density of baryons $n_{b}$ (in fm$^{-3}$).
The distances $R$ and $\Delta r$ (in km) are measured from
the center and surface of the star, respectively. We have set
the density at which the phase with nuclei changes into the liquid 
phase as $\approx 0.1$ fm$^{-3}$; the depth $\Delta r$ at this 
phase boundary is represented by the dotted line 
$\Delta r=r_{2}$, where $r_{2}=700$ m. 
The dotted line $\Delta r=r_{1}$, where $r_{1}=450$ m,
corresponds to the depth at a density $n_{b}\approx 0.02$ fm$^{-3}$.
\vspace*{0.5cm}

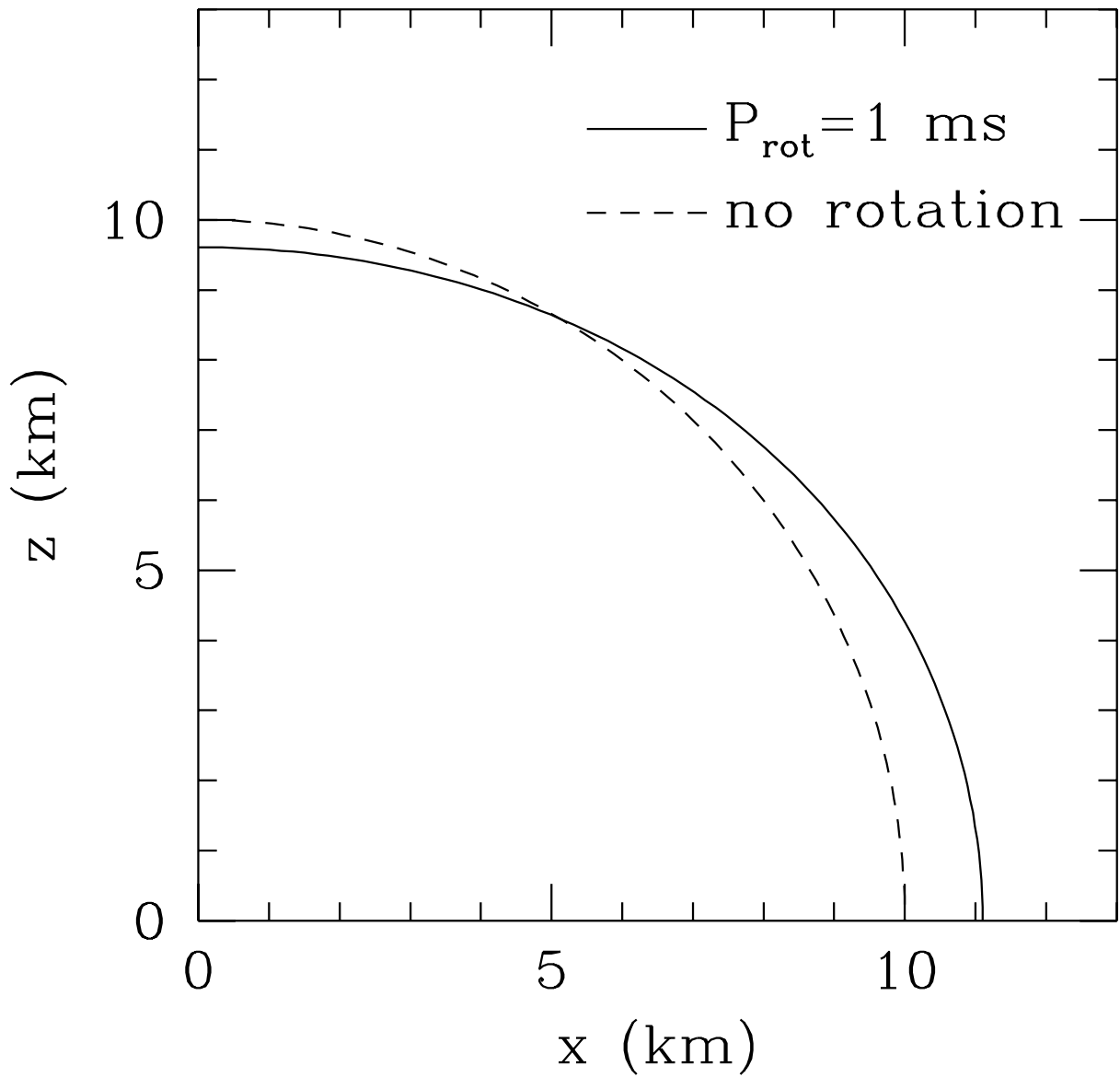
FIG. 4.-----  Surfaces of a neutron star in the rotating and
non-rotating configurations at fixed total baryon number. 
The solid curve is the surface of the star rotating at a rotation
period of 1 ms, and the dashed curve is the surface of the 
non-rotating star of radius 10 km.
The $z$ axis is taken to be the rotation axis.

\vspace*{0.5cm}

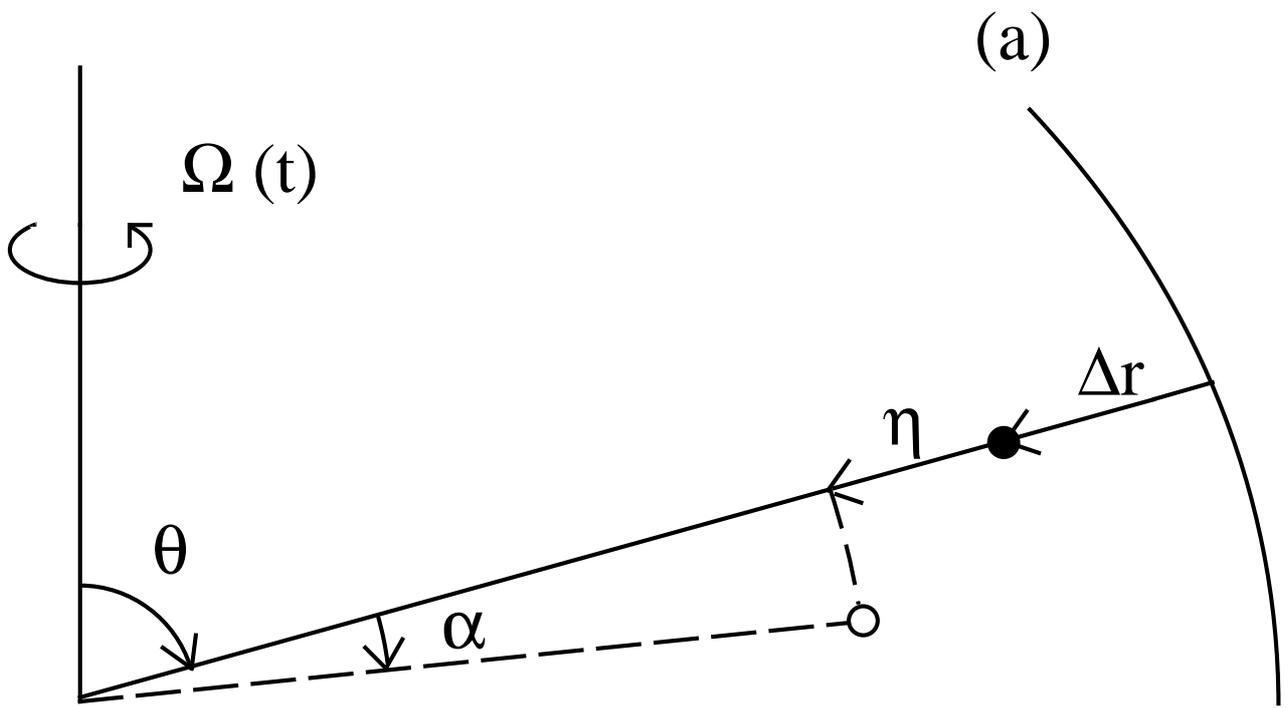
FIG. 5.-----  Definition of the quantities $\eta$ and $\alpha$, 
denoting the position of an element of matter in the crust 
at an age $t$ (open circle).
Its initial position (solid circle)
is denoted by $\Delta r$ 
and $\theta$. The surface (a) is the surface of a neutron star
rotating at an angular velocity $\Omega(t)$.

\vspace*{0.5cm}

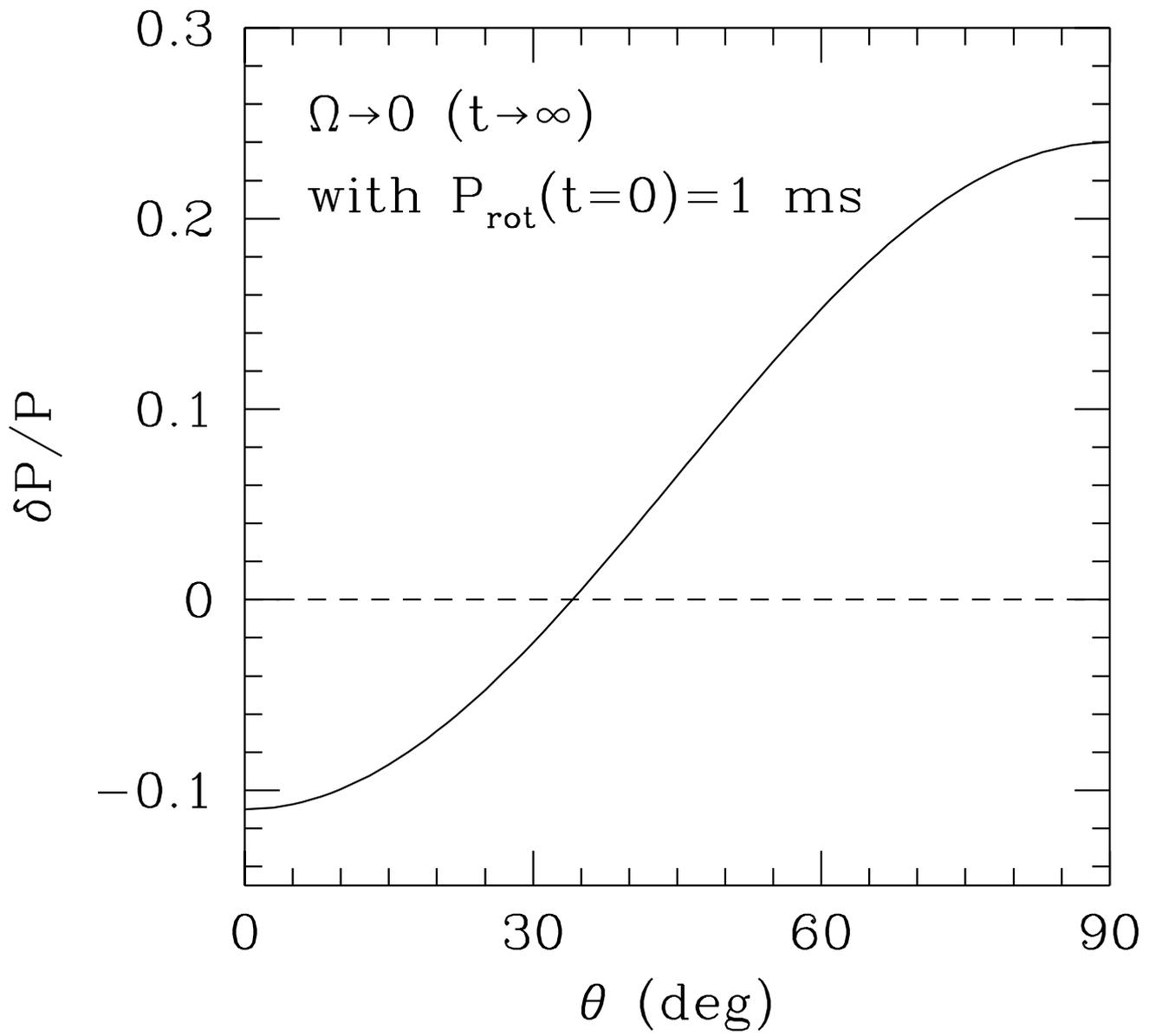
FIG. 6.-----  The change in the pressure 
of an element of matter in the crust
at an age when the rotation decays completely,
being divided by its initial pressure,
as a function of the polar angle
denoting its position.
The initial rotation period $P_{\rm rot}(t=0)$ 
is taken to be 1 ms.

\vspace*{0.5cm}

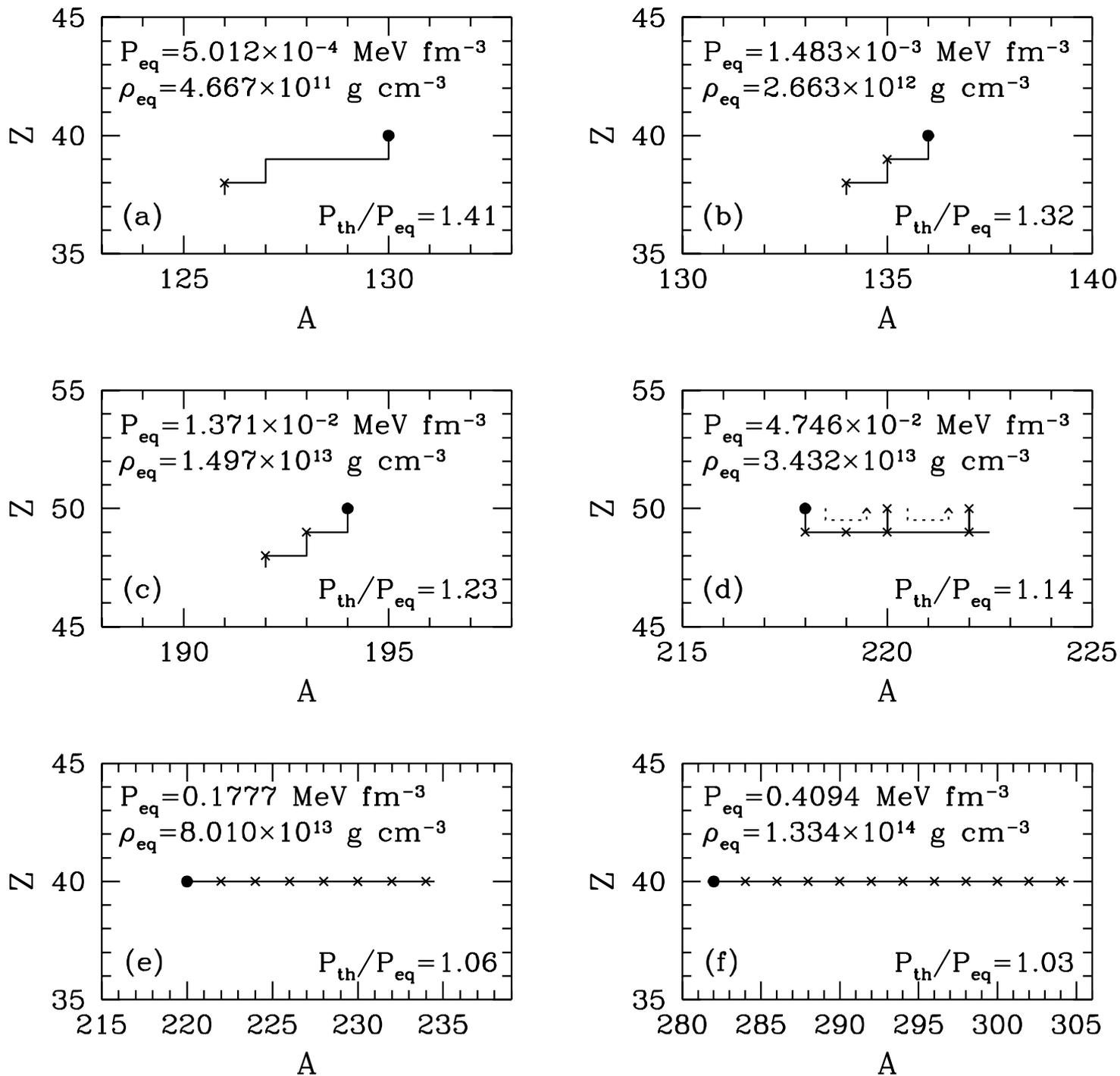
FIG. 7.-----  Changes in nuclides due to increase in pressure
at initial densities above neutron drip.
The solid circles are
the initial equilibrium nuclides 
at a density $\rho_{\rm eq}$ 
and a pressure $P_{\rm eq}$. The crosses are the nuclides in the
matter being in quasi-equilibrium.
The pressures $P_{\rm th}$ are the threshold
ones for the reactions occurring first.

\vspace*{0.5cm}

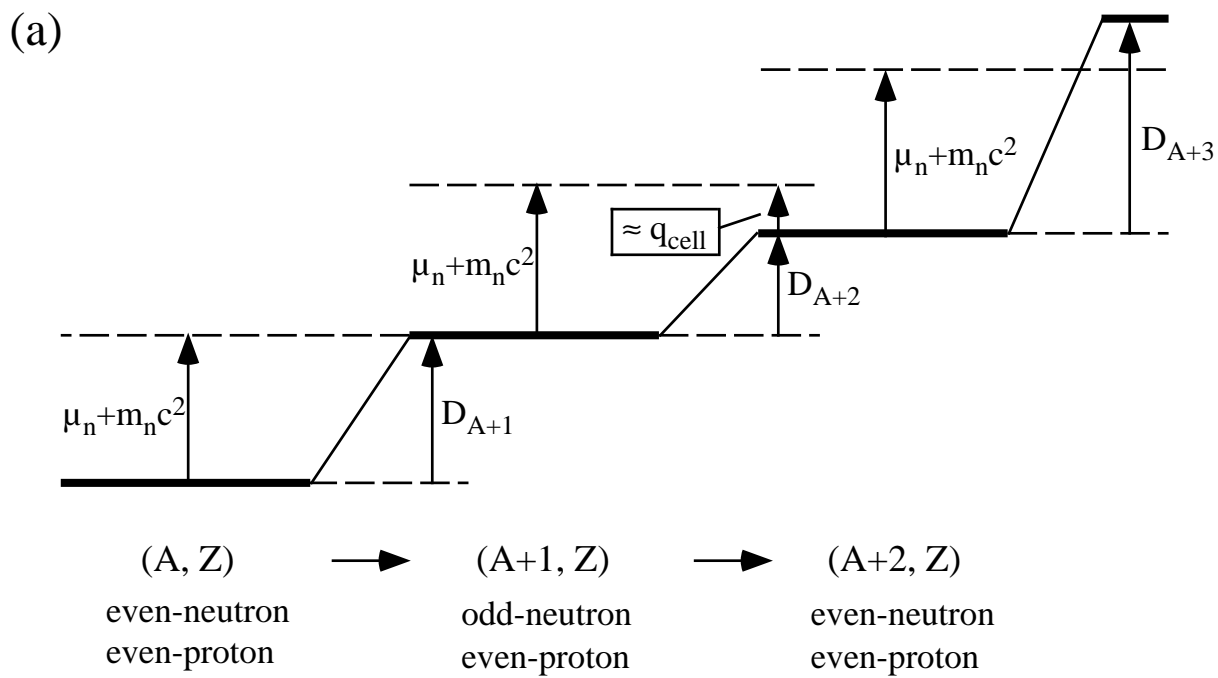
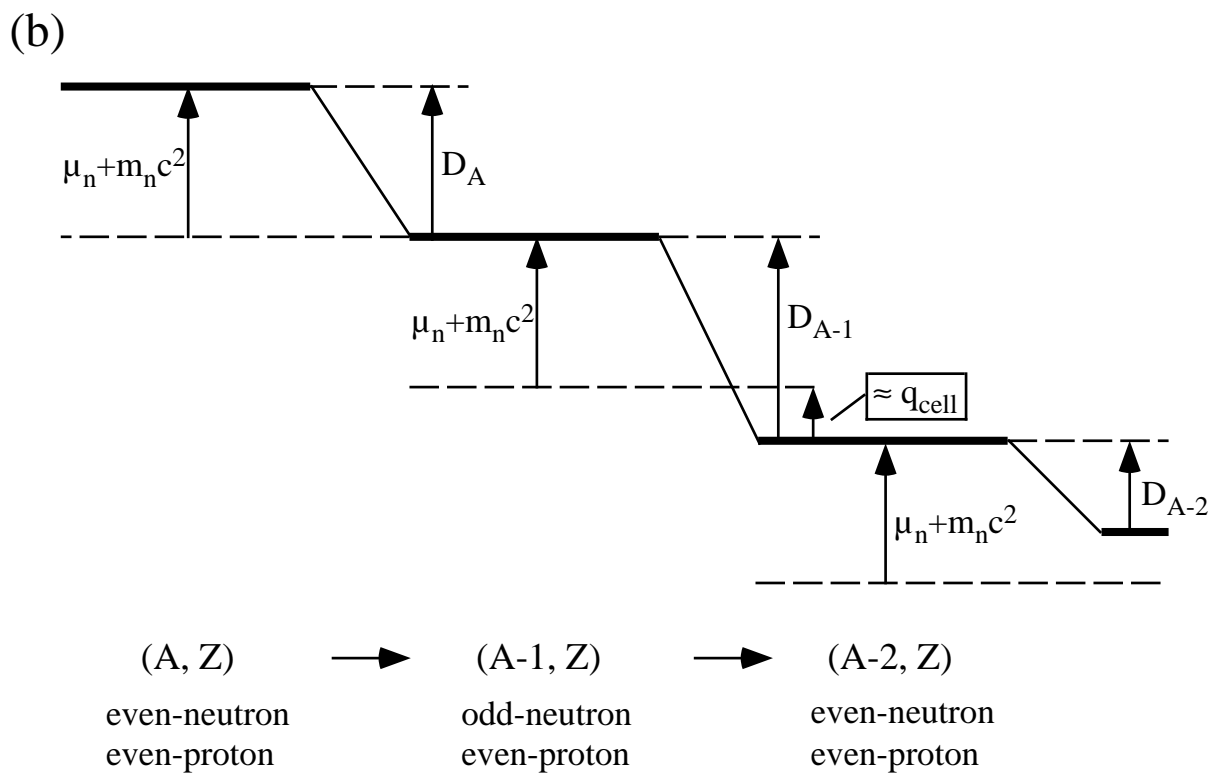
FIG. 8.-----  Schematic diagrams of nuclear levels relevant to
two-stage neutron (a) absorption and (b) emission, which take 
place in the matter initially present above a density of about
$3\times 10^{13}$ g cm$^{-3}$ during the compression and 
decompression, respectively, due to the spin-down with 
$P_{\rm rot}(t=0)=1$ ms. When the Fermi energy of dripped 
neutrons including the rest mass, $\mu_{n}+m_{n}c^{2}$, crosses
the energy threshold, $D_{A+1}$ $(D_{A})$, for neutron absorption
(emission) by the even-even nucleus $(A,Z)$, the nucleus changes
into the next one $(A+1,Z)$ $[(A-1,Z)]$ in a quasi-equilibrium
condition. Successively, another neutron absorption (emission)
occurs, which in turn releases the energy $qA_{\rm cell}$ in the
medium. The resulting state in which the system contains the even-even
nucleus stays in quasi-equilibrium. The two-step feature of the
reactions described above is attributable to the differences between
the values of $D_{A}$ for odd $A$ and those for even $A$, which
occur due to the nuclear pairing effects.

\vspace*{0.5cm}

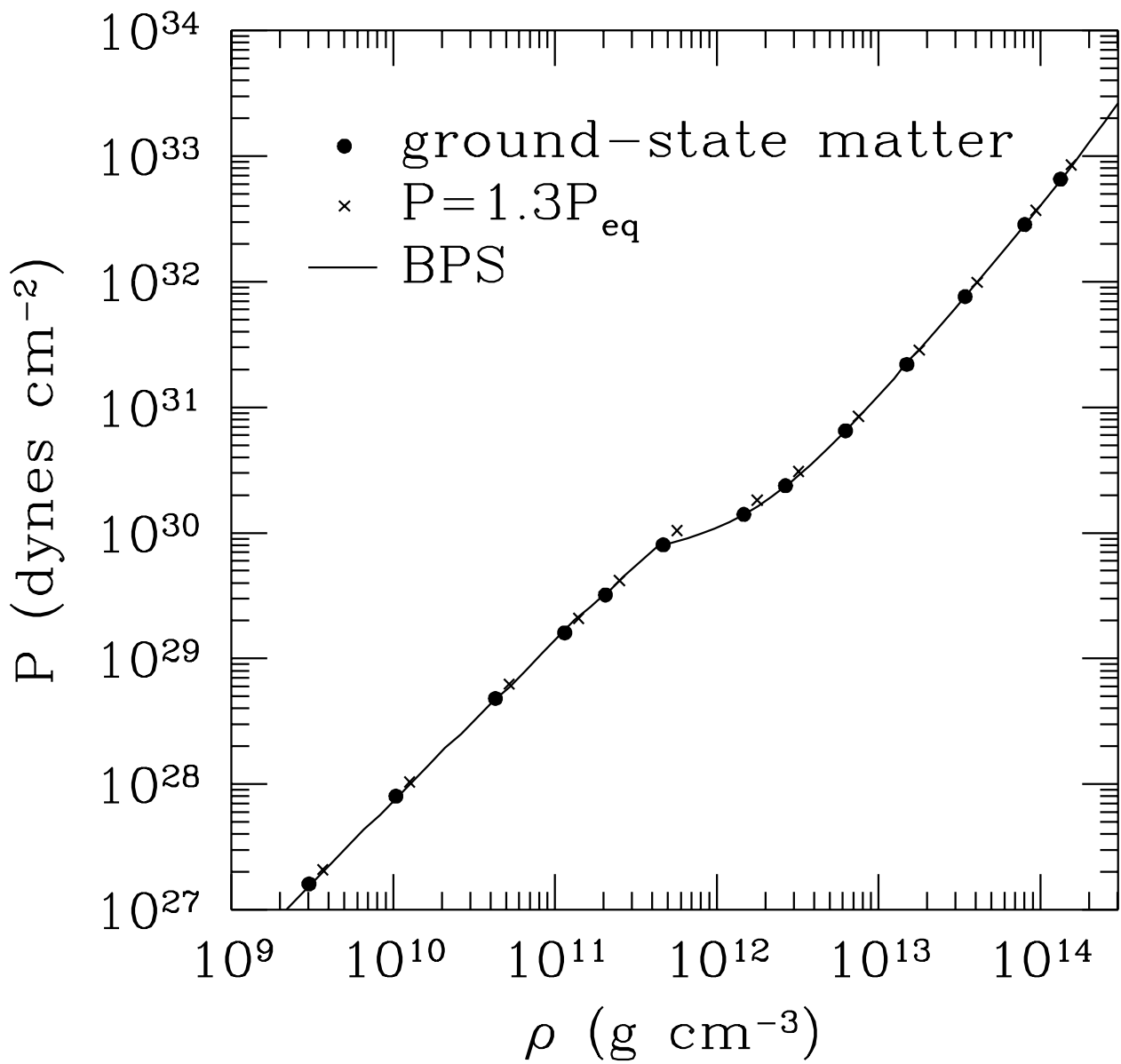
FIG. 9.-----  Changes in the equation of state of zero-temperature
matter in the crust from the ground-state one due to the compression.
The solid circles show the 
equilibrium equation of state, and
the crosses denote the equation of state of the matter 
deviated from nuclear equilibrium by an increase of 
30\% in its pressure. 
The solid line is the equation 
of state of BPS\@.

\vspace*{0.5cm}

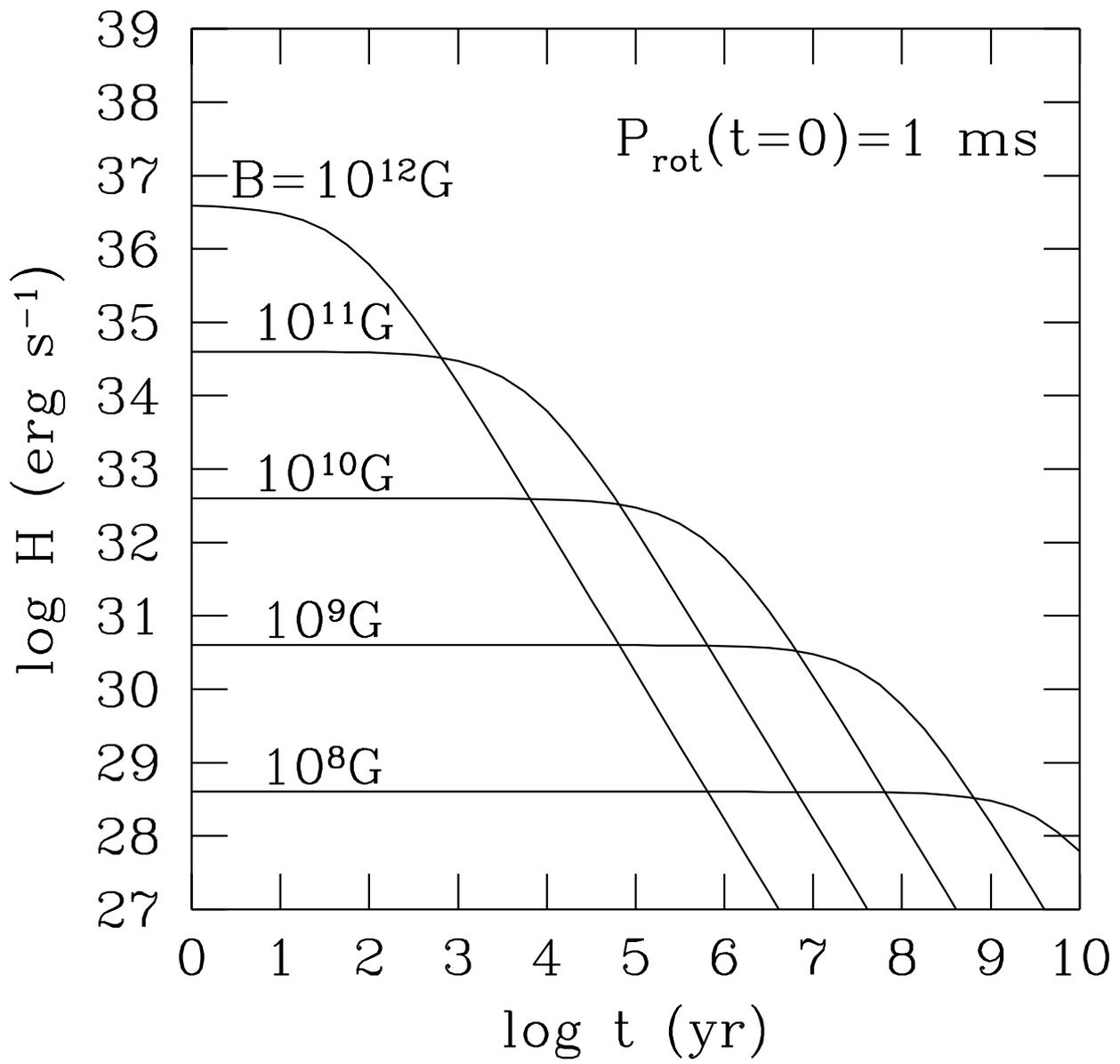
FIG. 10.-----  Heating rates as a function of the age of neutron stars
with various values of surface dipole magnetic fields.
The initial rotation period $P_{\rm rot}(t=0)$ 
is taken to be 1 ms.

\vspace*{0.5cm}

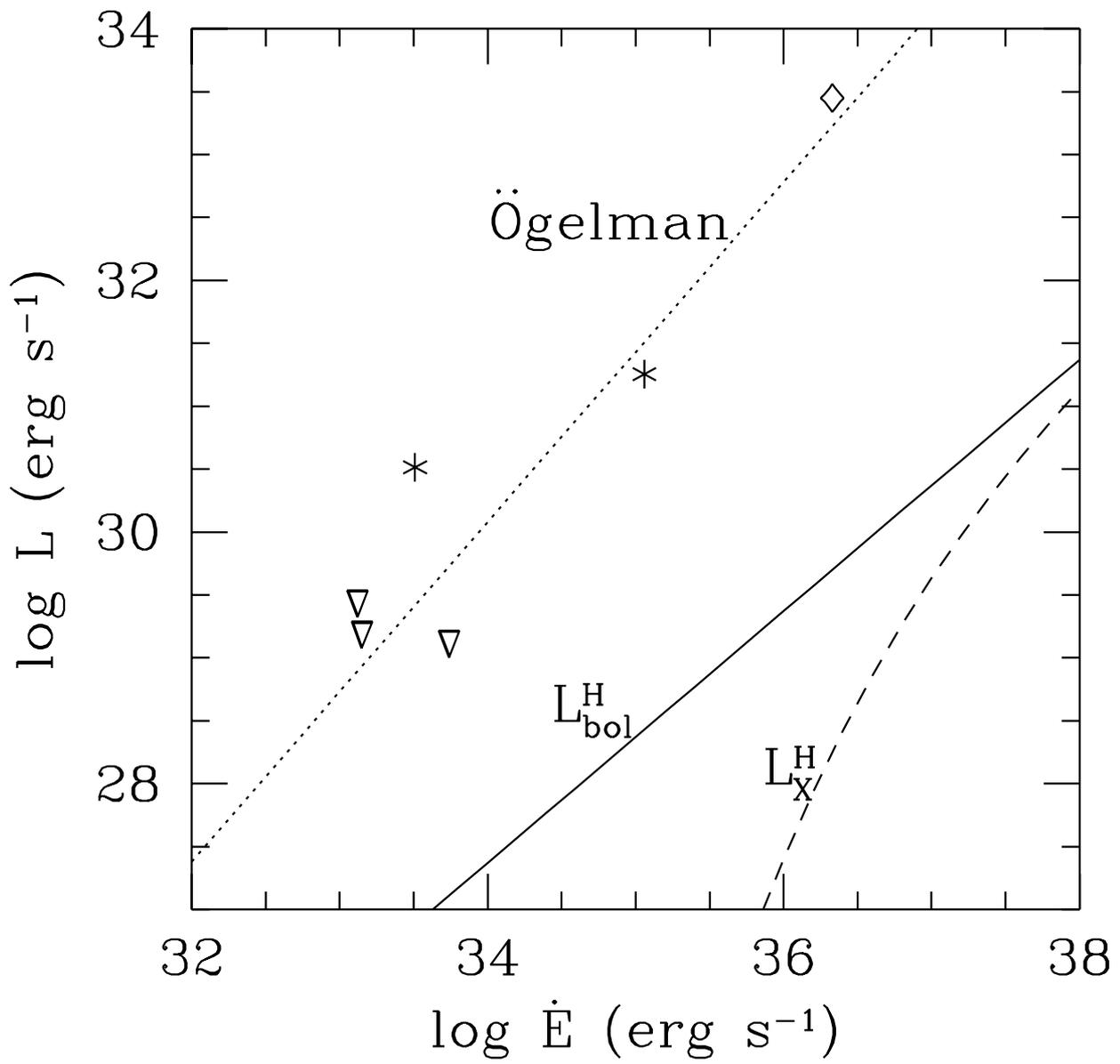
FIG. 11.-----  Photon luminosity from the surface of a neutron star
as a function of the spin-down power. The solid line is the bolometric
luminosity, $L_{\rm bol}^{H}$, predicted from the heating rate $H(t)$,
and the dashed line is the X-ray (0.1--2.4 keV) luminosity, 
$L_{X}^{H}$, predicted from $H(t)$ by assuming a blackbody spectrum. 
For comparison, the X-ray luminosities 
observed from some millisecond pulsars
(Danner et al.\ 1994) are shown. Triangles denote upper limits;
asterisks detections; a diamond a possible detection.  
The empirical curve (\"{O}gelman 1995; dotted line) for X-ray
luminosities coming from the stellar magnetosphere is also plotted.

\end{document}